\documentclass[reprint, amsmath,amssymb, aps,prb]{revtex4-1}
\usepackage{stackengine}
\usepackage{booktabs}
\usepackage{float}
\usepackage{ulem}

\usepackage[caption=false]{subfig}
\usepackage{graphicx}
\usepackage{bm}
\usepackage{hyperref}
\usepackage{natbib}
\usepackage[usenames,dvipsnames]{color}
\usepackage[usenames,dvipsnames,svgnames,table]{xcolor}
\usepackage{physics}

\usepackage{comment}
\bibpunct{\color{blue}[}{\color{blue}]}{,}{n}{}{;}
\hypersetup{
  colorlinks,
  citecolor=blue,
  linkcolor=blue,
  urlcolor=blue
  }

\newlength{\mysize}

\begin{document}
\preprint{APS/123-QED}
\title{Prevalence of trivial zero-energy sub-gap states in non-uniform\\
helical spin chains on the surface of superconductors}
\author{Richard Hess}
\author{Henry F. Legg}
\author{Daniel Loss}
\author{Jelena Klinovaja}%
\affiliation{%
 Department of Physics, University of Basel, Klingelbergstrasse 82, CH-4056 Basel, Switzerland }%
 \date{\today}
 \begin{abstract}
Helical spin chains, consisting of magnetic (ad-)atoms, on the surface of bulk superconductors are predicted to host Majorana bound states (MBSs) at the ends of the chain.  Here, we investigate the prevalence of trivial zero-energy bound states in these helical spin chain systems. The existence of trivial zero-energy bound states can prevent the conclusive identification of MBSs and, given the limited tunability of atomic spin chain systems, could present a major experimental roadblock. First, we show that the Hamiltonian of a helical spin chain on a superconductor can be mapped to an effective Hamiltonian reminiscent of a semiconductor nanowire with strong Rashba spin-orbit coupling. In particular, we show that a varying rotation rate between neighbouring magnetic moments maps to smooth non-uniform potentials in the effective nanowire Hamiltonian. Previously it has been found that trivial zero-energy states are abundant in nanowire systems with  smooth potentials. Therefore, we perform an extensive search for zero-energy bound states in helical spin chain systems with varying rotation rates. Although bound states with near zero-energy do exist for certain dimensionalities and rotation profiles, we find that zero-energy bound states are far less prevalent than in semiconductor nanowire systems with equivalent non-uniformities.  In particular, utilising  varying rotation rates, we do not find zero-energy bound states in the most experimentally relevant setup consisting of a one-dimensional helical spin chain on the surface of a three-dimensional superconductor, even for profiles that produce near zero-energy states in equivalent one- and two-dimensional systems. Although our findings do not rule them out, the much reduced prevalence of zero-energy bound states in long non-uniform helical spin chains compared with equivalent semiconductor nanowires, as well as the ability to measure states locally via STM, should reduce the experimental barrier to identifying MBSs in such systems. 
\end{abstract}

\maketitle

\section{Introduction \label{Sec:Introduction}}
The experimental realisation of Majorana bound states (MBSs) has become one of the most sought-after goals in modern condensed matter physics. The search for MBSs has largely been motivated by their exotic non-Abelian braiding statistics, which makes them a promising basis for fault tolerant quantum computation \cite{Volovik1999Fermion,Read2000Paired,Senthil2000Quasiparticle,Ivanov2001NonAbelian,Nayak2008Non,alicea2012new}. Despite this intense effort, however, there has been no conclusive observation of MBSs so far.

Topological $p_x+i p_y$ superconductors are predicted to host MBSs at the cores of vortices \cite{Volovik1999Fermion,Read2000Paired}. However, since intrinsic $p$-wave superconductors turn out to be rare, the main experimental focus has been on engineering hybrid platforms based on proximity effect that can become topological superconductors. A wide variety of engineered systems have been proposed to host MBSs \cite{alicea2012new,beenakker2013search,Pawlak2019Majorana, Laubscher2021Majorana,Flensberg2021Engineered} such as edge or surface states of topological insulators (TIs) 
 \cite{Fu2008}, semiconductor nanowires \cite{Sato2010NonAbelian,OregHelical2010,LutchynMajorana2010,Gangadharaiah2011Majorana,Sticlet2012Spin, Dominguez2012Dynamical, Prada2012Transport}, 
planar Josephson junctions \cite{BlackSchaffer2011Majorana,Pientka2017,Volpez2020Time}, TI nanowires \cite{Cook2011,Legg2021}, graphene-based systems \cite{Klinovaja2012, Sau2013, Klinovaja2012, Klinovaja2012Helical, BlackSchaffer2012Edge, Klinovaja2013Giant, Klinovaja2013Spintronics, Dutreix2014Majorana, SanJose2015Majorana, Zhou2016Ising, Dutreix2017Topological}, 
and many more. 

 \begin{figure}[t]
\subfloat{\stackinset{l}{0.0in}{t}{1pt}{{\color{black}(a)}}{\stackinset{l}{1.62in}{t}{0pt}{}{\includegraphics[width=1\columnwidth]{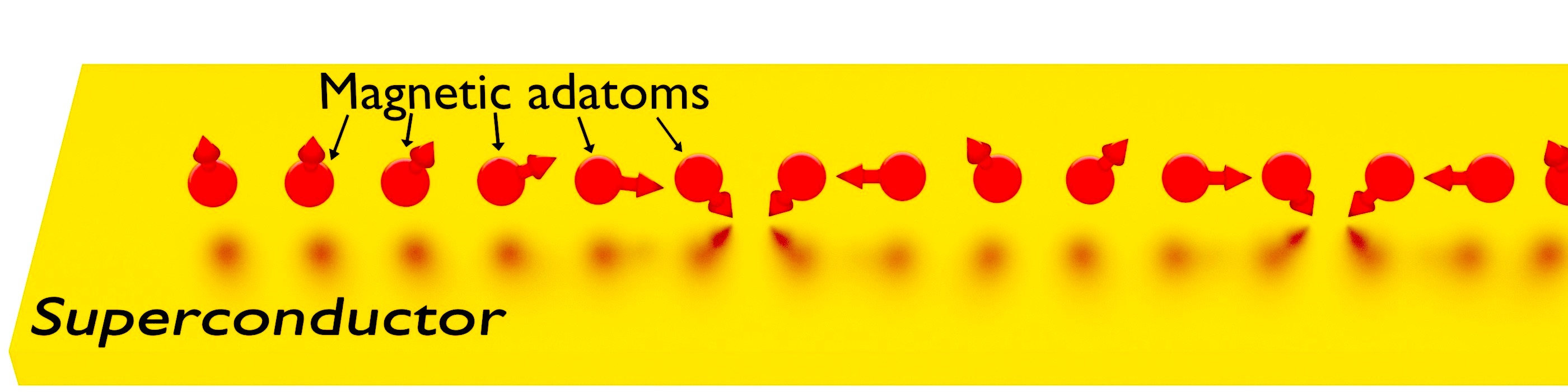}}}
}\\
\subfloat{\label{fig1dSchematicsChain}\stackinset{l}{0.00in}{t}{-1pt}{(b)}{\stackinset{l}{1.62in}{t}{0.in}{}{\includegraphics[width=1\columnwidth]{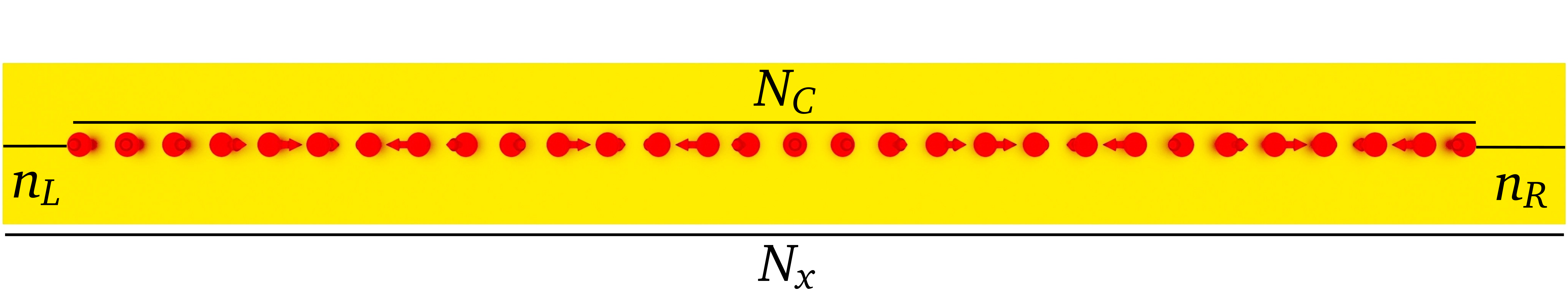}}}
}
 \caption{\textit{Helical spin chain system:} (a) Magnetic adatoms (red) are placed on the surface of a bulk superconductor (yellow). The adatoms form a helical spin chain with a rotation rate that can vary along the chain. (b) Geometry of helical spin chain used in this manuscript (number of sites shown is not to scale). The full system has $N_x$ sites in the $x$ direction with magnetic adatoms deposited on $N_C$ sites forming a helical spin chain. In general, the helical spin chain is embedded in the underlying superconductor with $n_L$ sites on the left side of the system and $n_R$ on the right. In addition  we will consider one-, two-, and three-dimensional superconductors underlying the chain.}
 \label{figPictureHelicalSpinChain2}
\end{figure}

Although many systems have been predicted to realise MBSs, to date, the conclusive experimental identification of MBSs has not been possible,  largely due to trivial states that can mimic the experimental signatures of MBSs \cite{Prada2020From}. Most notable is the situation in semiconductor nanowires, which are perhaps the most mature experimental platform expected to host MBSs. Such systems consist of a nanowire with strong Rashba spin-orbit coupling that has been brought into proximity with a superconductor.  It was predicted that a signature of MBSs in these nanowires is a zero-bias peak in differential-conductance measurements and that this peak is stable for a wide range of magnetic field strengths. Although such a signature has been observed
 \cite{Sasaki2011Topological,MourikSignatures2012,Deng2012Anomalous,Das2012Zero, Churchill2013Superconductor,Ruby2015EndStates,AlbrechtExponential2016, Pawlak2016Probing, schneider2021controlled}, it turned out that a zero-bias peak by itself is not a unique fingerprint of MBSs. In fact, it was shown that trivial Andreev bound states (ABSs) \cite{Andreev1964Thermal,Andreev1966Electron} are expected to be abundant in these nanowires \cite{Prada2020From}. These trivial states can also result in zero-bias peaks which mimic MBSs and they occur, for instance, due non-uniformities in the nanowire parameters \cite{kells2012Near,Lee2012Zero,Cayao2015SNS,Ptok2017Controlling,Liu2017Andreev, reeg2018zero,Penaranda2018Quantifying,moore2018two,Vuik2019Reproducing, Woods2019Zero,Liu2019Conductance,Chen2019Ubiquitous,Alspaugh2020Volkov, Juenger2020Magnetic,valentini2020nontopological,Hess2021Local,Marra2022Majorana, Roshni2022Conductance}. 
 Despite several further signatures of MBSs being proposed, due to the high prevalence of possible zero-energy modes, it remains unclear if a conclusive measurement of MBSs can be performed in nanowire systems.

Artificial magnetic structures, such as atomic chains, where adatoms are placed on the surface of a superconductor have also been predicted to host MBSs  
\cite{Klinovaja2013TopologicalSuper,Braunecker2013Interplay,Vazifeh2013Self,Heimes2014Majorana,Hui2015Majorana, Peng2015Strong,Heimes2015Interplay,Choy2011Majorana,Martin2012Majorana,Perge2013Proposal,Pientka2013Topological,Pientka2014Unconventional, Reis2014SelfOrganized,Poeyhoenen2014Majorana}. 
A single magnetic adatom on the surface of a superconductor leads to the formation of a Yu-Shiba-Rusinov (YSR) state, which is well studied in theory \cite{Yu1965YSR, Shiba1968Classical,Rusinov1969On,Flatte1997Local,Flatte1997Defects,Salkola1997Spectral,Balatsky2006Impurity} and experiment \cite{Ali1997Probing,Ji2008High,Perge2014Observation,Ruby2015Tunneling, Menard2015Coherent,Ruby2018WaveFunction,kuster2021Correlating, Kuster2021Long,Wang2021Spin,Flatte2000Local,Hoffman2015Impurity}.
Current experiments exploit an atom by atom construction technique to build the atomic chains,  allowing great control over the chain length  
 \cite{Feldman2017High,Sangjun2017Distinguishing,Steinbrecher2018Non, Jack2021Detecting, Howon2018Toward, Schneider2021Topological,Kuster2021Long,kuester2021nonmajorana, Ding2021Tuning,Schneider2022Precursors}.
 Here we will focus on long chains of magnetic adatoms that possess a helical ordering. Helical ordering of the magnetic moments of the adatoms can be mediated by Ruderman-Kittel-Kasuya-Yosida (RKKY) interaction \cite{Rudermann1954Indirect,Kasuya1956Theory,Yosida1957Magnetic,Braunecker2009Nuclear,Braunecker2010SpinSelective} which should lead to a helix of magnetic moments with period $\pi/k_F$, where $k_F$ denotes the Fermi momentum.  Other mechanisms can also support helical ordering, for instance the Dzyaloshinskii-Moriya interaction (DMI) \cite{Menzel2012Information}. Since local measurements can be performed on such atomic chains by utilising scanning-tunnelling microscopy techniques, the location of states can be very well established \cite{Howon2018Toward,schneider2021controlled,Schneider2021Topological}, a significant benefit over semiconductor systems. On the other hand, the topological transition in such chains is set by the exchange coupling strength, $J$, between the adatoms and the superconductor, which is not easily controlled. This lack of tunability makes exploring the phase space of a zero-energy bound state difficult and therefore, if trivial zero-energy states are as abundant in such chains as in nanowire systems, it would be even more difficult to conclusively identify MBSs.

In this paper we will investigate the prevalence of trivial zero-energy modes due to non-uniformities in long helical spin chains. In particular, we will consider chains with varying rotation rates between adjacent magnetic moments along the chain. We show that a mapping exists from the Hamiltonian of a helical spin chain to an effective Hamiltonian reminiscent of a Rashba nanowire. Under this mapping varying  rotation rates of the magnetic moments result in non-uniformities in the effective Hamiltonian that are similar to those that result in the abundance of trivial states in Rashba nanowires. Therefore, using this mapping as a basis, we will investigate if non-uniform rotation rates can also lead to an abundance of zero-energy modes in helical spin chains. Although bound states with near zero-energy do exist for certain dimensionalities and rotation profiles, including when the spin chain remains entirely trivial, for experimentally relevant systems we find that zero-energy bound states are far less prevalent in spin chains than semiconductor nanowire systems. Primarily this is due to the fact the helical chain is embedded in a superconductor and also due to effects of the boundary.

This work is structured as follows: first, we investigate an effective one-dimensional continuum model for a helical spin chain with a smooth spatially varying rotation rate between adjacent magnetic moments on top of a superconductor, showing that this maps to a Hamiltonian similar to that of a Rashba nanowire, see Sec.~\ref{Sec:ModelNanowires}. In Sec.~\ref{Sec:LatticeModels} we introduce one-, two- and three-dimensional lattice models for the non-uniform helical spin chains. The topological phase diagram of a helical periodic one-dimensional spin chain is shown in Sec.~\ref{Sec:PhaseDiagram} and, in addition, several non-uniform profiles of the rotation rate are suggested, which might potentially support the formation of trivial zero-energy sub-gap states. In Secs.~\ref{Sec:SmoothDecay}, \ref{Sec:DomainWall}, and \ref{Sec:QuasiMBS}, we discuss the presence and absence of trivial zero-energy sub-gap states in one-, two- and three-dimensional systems for different types of rotation rate profiles. Finally, we conclude in Sec.~\ref{Sec:Conclusion}. In addition, in App.~\ref{App:2dChainOnEdge}, we study the disappearance of trivial zero-energy states in chains which are placed on the boundary of a two-dimensional superconductor. Furthermore, we consider the scenario of a chain  deposited in the bulk  of a three-dimensional superconductor  instead on the surface, see App.~\ref{App:ChainIn3dBulk}.

\section{Theory of a one-dimensional spin chain with spatially varying magnetization \label{Sec:ModelNanowires}  \label{Sec:Model}}

In this section we show that the Hamiltonian of a helical spin chain, as shown in Fig.~\ref{figPictureHelicalSpinChain2}, can be mapped to an effective Hamiltonian that is reminiscent of a semiconductor nanowire brought into proximity with a superconductor. Surprisingly, despite this mapping and despite the fact that spatially varying potentials can easily be generated in the effective Hamiltonian, we will show in the remainder of the manuscript that  trivial zero-energy bound states are far less abundant in helical spin chains than has been shown to occur in semiconductor Rashba nanowires.

To begin, we consider a helical spin chain placed on top of a superconductor. The effective one-dimensional (1D) model is described by the Hamiltonian  $\mathrm{H}_{\mathrm{1D}}=\int \dd x\ \Psi^{\dagger}(x) H_{\mathrm{1D}} \Psi(x)$  in the basis 
\begin{align}
\Psi(x)=\begin{pmatrix}
c_{x,\uparrow}, & c_{x,\downarrow}, & c^{\dagger}_{x,\downarrow}, & -c^{\dagger}_{x,\uparrow} \\
\end{pmatrix}^{\rm T},
\end{align}
where the operator $ c_{x,\nu}^{\dagger} $ ($c_{x,\nu}$) creates (annihilates) an electron at  position $x$ with spin $ \nu $. The Hamiltonian density is given by 
\begin{align}
H_{\mathrm{1D}}= \left(-\frac{\hbar^2}{2m}\nabla_x^2-\mu\right)\tau_z+J(x)\mathbf{S}(x)\cdot\boldsymbol{\sigma}+\Delta_0 \tau_x \label{eq:effHam},
\end{align}
where $m$ is the effective mass, $\mu$ the chemical potential, $\Delta_0$ is the superconducting gap, and $J(x)$ is the (position dependent) exchange coupling strength between the magnetic moments of the adatoms and the spins of the itinerant electrons. In addition, $\boldsymbol{\sigma}=(\sigma_x,\sigma_y,\sigma_z)^{\rm T}$ is the vector of Pauli matrices acting in spin space and $\tau_i$ are Pauli matrices acting in particle-hole space. Finally, the unit-vector $\mathbf{S}(x)$ determines the local direction of the magnetic moments of the adatoms that form the chain. We will consider magnetic textures along the chain of the form
\begin{align}
\mathbf{S}(x)=\begin{pmatrix}
\cos[2\varphi(x)],& \sin[2\varphi(x)], &0
\end{pmatrix},\label{eq:SpinVector}
\end{align}
such that the rotation of the magnetic moments in the $xy$-plane is described by the total angle  $\varphi(x)$ which can be described by a non-linear function, assuming a non-uniform rotation. Applying the unitary transformation $U=e^{-i\varphi(x)\sigma_z}$, one can map the Hamiltonian of the helical chain, $H_{\mathrm{1D}}$,  to that of a Rashba nanowire in a magnetic field $\tilde{H}$ \cite{Braunecker2010SpinSelective}, such that
\begin{align}
\tilde{H}=&U^{\dagger} H_{\mathrm{1D}}
 U\nonumber \\
=&-\frac{\hbar^2}{2m} \left(\nabla_x^2-\left(\!\pdv{\varphi}{x}\!\right)^{\!2}-\left[i\pdv{\varphi}{x}\nabla_x +i\nabla_x \pdv{\varphi}{x}\right]\sigma_z\right)\tau_z\nonumber \\
&+ J(x)\sigma_x+\Delta_0 \tau_x - \mu\tau_z\label{eq:TransformedHelicalChain}.
\end{align}
When the rotation rate of the magnetic moments along the chain, which we define as $\Phi(x)\equiv \pdv{\varphi}{x}$, is non-uniform then the transformed Hamiltonian, $\tilde{H}$, contains an effective coordinate dependent potential $\tilde{V}(x)=\hbar^2 [\Phi(x)]^2 / 2m$  and also an effective coordinate dependent Rashba SOI
\begin{align}
\tilde{H}_R(x)&=\frac{\hbar^2}{2m}[i\Phi(x)\nabla_x +i\nabla_x \Phi(x)]\sigma_z\tau_z\nonumber \\
&=\frac{i}{2}\left[\alpha(x)\nabla_x+\nabla_x\alpha(x)\right]\sigma_z\tau_z,
\end{align}
where we use the symmetrized from of the effective position-dependent Rashba SOI \cite{Klinovaja2015Fermionic} with the strength given by  $\alpha(x)= \hbar^2 \Phi(x) / m$.  The same unitary transformation $U$ will also yield an equivalent Hamiltonian in the generalisation of our system to two- or three-dimensions. This is because the kinetic terms that contain the derivative $\nabla_i$, acting in the direction $i \neq x$ with $i  \in\lbrace y,z\rbrace $, will commute with $U$.
 
The transformed Hamiltonian, $\tilde{H}$, is reminiscent of one-dimensional semiconductor nanowires and has the same ingredients that are required for topological superconductivity: Rashba SOI, a superconducting gap, and a Zeeman energy \cite{OregHelical2010,LutchynMajorana2010}. Using this mapping we therefore see that, in the case of a constant rotation rate between adjacent magnetic moments along the chain, which means $\varphi(x)=k_rx$, the topological phase transition for the purely one-dimensional chain takes place at
 \begin{align}
J_C(k_r)= \sqrt{\Delta_0^2+(\mu-\hbar^2k_r^2/[ 2m])^2} \label{eq:PhaseTransitionContinous}
\end{align} 
and depends on the spatial rotation period, which is set by $2\pi/(2k_r)$. If the magnetic moments of the magnetic adatoms are aligned by RKKY interaction ($k_r=k_F$), then the period is set to $\pi/k_r=\pi/k_F$ and the system enters the topological phase at $J(k_F)=\Delta_0$ \cite{Klinovaja2013TopologicalSuper,Braunecker2013Interplay,Vazifeh2013Self}. The topological phase transition is shifted to larger values of $J$ for $k_r\neq k_F$. 

Therefore, when the rotation rate of the one-dimensional helical chain is non-uniform, the above mapping is to an effective Hamiltonian that is reminiscent of a one-dimensional semiconductor nanowire with spatially varying parameters. Such nanowires have been considered extensively and it has been shown that trivial zero-energy Andreev bound states (ABSs) are abundant in these systems \cite{kells2012Near,Lee2012Zero,Cayao2015SNS,Ptok2017Controlling,Liu2017Andreev,
reeg2018zero,Penaranda2018Quantifying,moore2018two,Vuik2019Reproducing,Woods2019Zero,
Liu2019Conductance,Chen2019Ubiquitous,Alspaugh2020Volkov,Prada2020From,
Juenger2020Magnetic,valentini2020nontopological}. The existence of the mapping therefore suggests that trivial zero-energy states could potentially be as prevalent in helical spin chains as in nanowires, which would be a significant problem given the more limited tunability of parameters in the atomic chain compared to a nanowire. However, although the mapping above is rigorous, the two systems differ significantly for the following reason: superconductivity in a semiconductor nanowire is present only in the region that is brought into proximity with a superconductor and the Zeeman energy is (approximately) constant throughout this region. In contrast, the helical spin chain is embedded in a superconductor and the exchange coupling energy -- which maps to the Zeeman energy -- is non-zero only close to the position of the magnetic adatoms.  As a result,  as we will show in the rest of this paper, while trivial zero-energy modes in a purely one-dimensional chain that is aligned with the end of the superconductor are as easily generated as this mapping suggests, the same is {\it not} true when the atomic chain is embedded in a superconductor, especially in the experimentally relevant scenario where the superconductor in which the chain is embedded is three-dimensional.

\section{Lattice models of spin chain \label{Sec:LatticeModels} }
\subsection{One-dimensional system\label{Sec:ModelOneDimensionDiscretized}}

In the previous section, we work in the continuum limit. In this section, we switch to the lattice description by discretizing the Hamiltonian, which will allow us later to solve the problem numerically. First, we construct a lattice Hamiltonian $H_{1\mathrm{D},L}$, corresponding to $H_{1\mathrm{D}}$ defined in Eq.~\eqref{eq:effHam}.
The lattice Hamiltonian  $H_{1\mathrm{D},L}$ of the setup consisting of an atomic chain on the superconducting surface (see Fig.~\ref{figPictureHelicalSpinChain2}) is given by 
\begin{align}
&H_{1\mathrm{D},L}=\sum_{n=1}^{N_x} \Bigg[\sum_{\nu,\nu'}c_{n,\nu}^{\dagger}( \lbrace 2t-\mu\rbrace\delta_{\nu,\nu'}+J_n\left(\mathbf{S}_n\cdot\boldsymbol{\sigma}\right)_{\nu \nu'}) c_{n,\nu'}\nonumber \\
&\hspace{30pt}-\Bigg(\sum_{\nu}t c_{n,\nu}^{\dagger}  c_{n+1,\nu}+\Delta_0 c_{n,\downarrow}^{\dagger}c_{n,\uparrow}^{\dagger}  +\text{H.c.} \Bigg) \big. \Bigg] \label{eq:effHamDiscretized},
\end{align}
where $N_x$ denotes the total number of lattice sites and $t=\hbar^2/(2ma^2)$ denotes the matrix hopping element, which depends on the effective lattice constant $a$.  We consider an exchange coupling strength of the form
\begin{align}
J_n=J\left[\Theta(n-n_{L})-\Theta(n-N_x+n_R)\right],
\end{align}
where $n_{L}$ ($N_x-n_R$) is the site hosting the first (last) adatom and $\Theta(n)$ is the Heaviside step-function with $\Theta(0)=0$ and the length of the chain of magnetic adatoms is given by $N_C=N_x-(n_{R}+n_{L})$. Depending on the relative locations of the ends of the chain and the system boundaries, we distinguish two different types of setups:  1) An \textit{aligned} setup is defined as the following: an end of the chain coincides with  a boundary of the system, such that $N_C=N_x$, [see Fig.~\ref{figPictureHelicalSpinChain2}(b)]. 2)~In an \textit{embedded} setup,  in contrast, magnetic adatoms are located only along a subsection of the entire system and do not reach the system boundary. In particular, we will consider an embedded left end, such that $N_x\gg N_C$ and  $n_R=0$ [see Fig.~\ref{figPictureHelicalSpinChain2}(b)]. We will use the same nomenclature of aligned and embedded systems referring to setups in which the superconductor underlying the chain is two- or three-dimensional (see below).

We define the rotation rate, $\Phi_n$, such that the helix has a period $\pi a /\Phi_n$, as
\begin{align}
\Phi_n(\lambda)=\Phi_L+(\Phi_R-\Phi_L)\ \text{sig}\left(\frac{2[n-n_0]}{\lambda}\right) \label{eq:AngleOmega},
\end{align}
where we have used the sigmoid function $\text{sig}(x)=\frac{1}{2}\left[1+\tanh(x/2)\right]$.
The profile $\Phi_n$ describes a  helix of magnetic moments which rotates with the period $\pi a/\Phi_{L}$ and $\pi a/\Phi_{R}$ on the left and right end of the chain, respectively. If the sign of $\Phi_{L}$ is the opposite of $\Phi_{R}$, then the rotation direction changes and the system contains a domain wall. The parameter $ \lambda $ controls the width and smoothness of the transition between two sections of different rotation rate and the parameter $ n_0 $ parameterises the position of the transition. In the discretized model $\varphi_n$ enters ${\bf S}(x)$ [see Eq.~\eqref{eq:SpinVector}] instead of $\varphi(x)$, where the angle $\varphi_n$ relative to the magnetic moment located at $n=1$ is given by
\begin{align}
\varphi_n=\sum_{m=1}^n  \Phi_{m}(\lambda)-\Phi_1.
\end{align} 
The particular choice of a constant rotation rate, $ \Phi_R=\Phi_L =\Phi$, results in the well known case of a helical spin chain with fixed rotation period ($\varphi_n=n \Phi-\Phi_1$). Consequently, the topological phase is defined by the condition \cite{Perge2013Proposal} 
\begin{align}
J^<_{C}(\Phi)=\sqrt{\Delta_0^2+(|\mu-2t|-2t|\cos(\Phi)|)^2}<J\nonumber \\<\sqrt{\Delta_0^2+(|\mu-2t|+2t|\cos(\Phi)|)^2}=J^{>}_C(\Phi),\label{eq:BothTPTs}
\end{align}
see Fig. \ref{figPhaseDiagramEnergyDifference}. In the limit of small angles between magnetic moments, $\Phi$, the prediction for the lower bound of the topological phase transition, $J_{C}^<$, corresponds to the analytic result, $J_C$, for the location of the bulk gap closing in the continuum system as discussed in the previous section [see Eq.~\eqref{eq:PhaseTransitionContinous}]. This can be shown explicitly by expanding the cosine function and using the definition of the matrix hopping element $t$ as well as the relation $\Phi=k_r a$. We note that the choice $\Phi_{k_F}=k_F a$ results in the topological phase transition criterion $J^<_{C}=\Delta_0$, which is the lowest value of $J^<_{C}$ possible for any $\Phi$. 
 
\subsection{Two- and three-dimensional systems \label{Sec:2dAnd3dModel}}
Here, we extend our model from Sec.~\ref{Sec:ModelOneDimensionDiscretized} to higher dimensions, such that the chain is deposited on top of a two- or three-dimensional (2D, 3D) superconductor. The Hamiltonian in this case has the form
\begin{align}
&H_{\kappa \mathrm{D}}=\sum_j\Bigg[ \sum_{\nu,\nu'}c_{j,\nu}^{\dagger}\lbrace (2t\kappa-\mu)\delta_{\nu,\nu'}+J_j (\mathbf{S}_{j}\cdot \boldsymbol{\sigma})_{\nu,\nu'}\rbrace c_{j,\nu'} \nonumber \\ 
&\hspace{20pt}-\Delta_0 c_{j,\downarrow}^{\dagger}c_{j,\uparrow}^{\dagger}-\Delta_0^* c_{j,\uparrow}c_{j,\downarrow}\Bigg]-\sum_{\langle j,j'\rangle,\nu} t c_{j,\nu}^{\dagger}c_{j',\nu} ,
\end{align}
where $\kappa \in \lbrace 2,3\rbrace$ denotes the dimensionality of the model and 
the index $j$ accounts for the $x$, $y$, and $z$ coordinates of the corresponding lattice site. In particular, we choose $j=(n,m)$ and  $j=(n,m,l)$ in two and three dimensions, respectively, and the number of sites in $y$ ($z$) direction is $N_y$ ($N_z$). The exchange coupling, for example in three dimensions, is given by $J_j=J_{n,m,l}=J_n\delta_{m,m_0}\delta_{l,l_0}$, where $m_0$ and $l_0$ denote the $y$ and $z$ position of the atomic chain, respectively.Here,  $J_n$ is the same as in the one-dimensional chain and the notation $\langle j,j' \rangle$ indicates a summation over nearest neighbouring sites. As discussed above, we also use the terms {\it aligned} and {\it embedded} to describe systems where $N_C=N_x$ and $N_C\ll N_x$, respectively. Finally we note that, since the atomic chain breaks translation symmetry in the directions perpendicular to the chain, momentum in these directions is not a good quantum number and the location of the topological phase transition can be expected to be different than that found in the equivalent purely one-dimensional system. We use the Python package Kwant for the implementation of the tight binding models \cite{Groth2014Kwant}.

\section{Topological phase diagram and rotation rate profiles\label{Sec:PhaseDiagram}}

\begin{figure*}[t]
\subfloat{\label{figPhaseDiagramEnergyDifference}\stackinset{l}{-0.00in}{t}{-0.in}{(a)}{\stackinset{l}{3.5in}{t}{-0.0in}{(b)}{\stackinset{l}{5.25in}{t}{-0.05in}{(c)}{\stackinset{l}{3.5in}{t}{1.12in}{(d)}{\stackinset{l}{5.25in}{t}{1.12in}{(e)}{\stackinset{l}{1.67in}{t}{1.76in}{}{\includegraphics[width=1\textwidth]{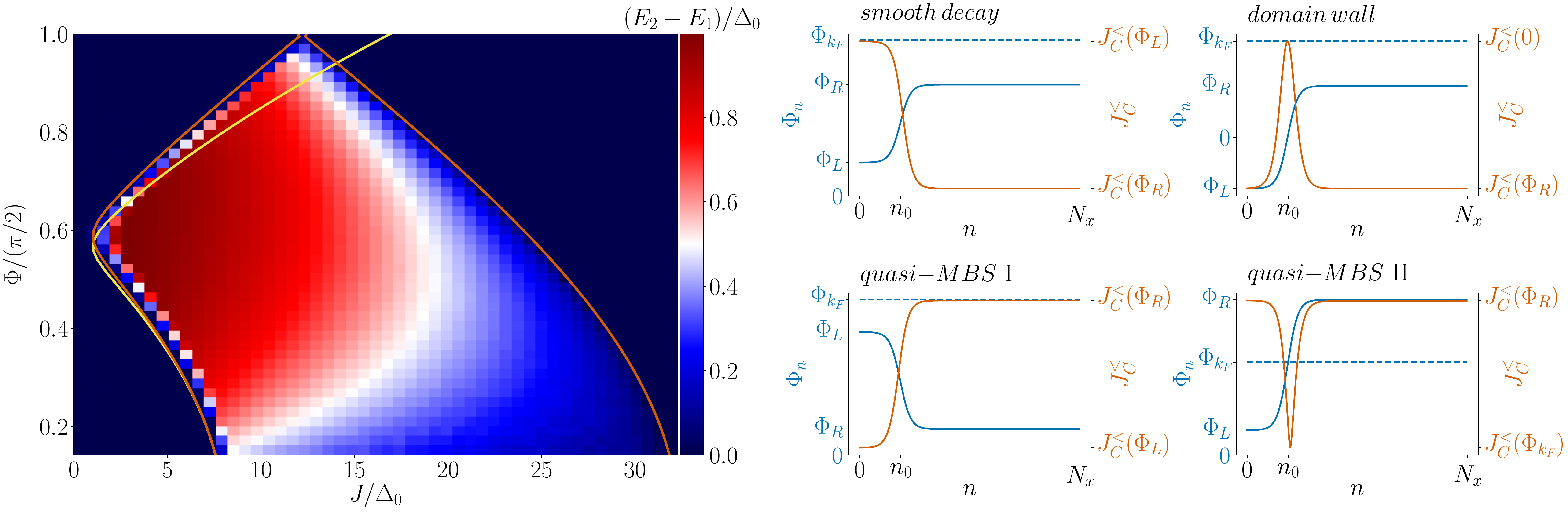}}}}}}}
}
\subfloat{\label{figAngleSmoothDecay}}
\subfloat{\label{figAngleDomainWall}}
\subfloat{\label{figAngleQuasiMBSFromBelowJT}}
\subfloat{\label{figAngleQuasiMBSCrossingkF}}
 \caption{\textit{Phase diagram and non-uniform rotation profiles:} (a): Phase diagram of a one-dimensional helical spin chain with constant rotation rate ($\Phi_n=\Phi=\mathrm{const.}$), showing the energy difference $ E_2-E_1$ between the lowest energy state and the first excited state. Orange lines indicate the critical exchange couplings $J^{<}_C(\Phi)$ and $J^{>}_C(\Phi)$, which separate the topological phase from the trivial phase [see Eq.~\eqref{eq:BothTPTs}], while the yellow line represents $J_C$ [see Eq.~\eqref{eq:PhaseTransitionContinous}]. In the trivial phase, this energy difference is given by the finite-size quantization and very small, corresponding to dark blue color.   For small rates $J_C$ agrees well with $J^{<}_C$. Next, we consider spatially varying rotation rates, with constant rates at the long right section of the chain:
  (b-e) Different profiles of the rate $\Phi_n$ (blue) and the corresponding profile of $J^{<}_C(\Phi_n)$ (orange) along the atomic chain to be considered in this work. The blue dashed line indicates the value of $\Phi_{k_F}$ corresponding to a period $\pi/k_F$. Numerical parameters for the phase diagram: $t\approx 10$ meV, $N_x=N_c=322$, $\Delta_0=1$ meV, $a=3 \, \mbox{\normalfont\AA}$, $\mu=8$ meV.}
 \label{figPhasediagramAndAngleConfigurationsAndExchangeCoupling}
\end{figure*}

In this section, we briefly discuss the topological phase diagram of the helical spin chain and suggest different rotation rate profiles that, based on the mapping to semiconductor nanowires (see Sec.~\ref{Sec:Model}), could lead to low energy trivial sub-gap states. First, we start with an aligned one-dimensional system and plot the energy difference $E_2-E_1$ between the first excited state  and the ground state  for a  chain with constant $\Phi_n \equiv \Phi$, such that sub-gap states other than MBSs or ABSs close to the gap edge, emerging for example from a shift of the chemical potential \cite{Huang2018Meta,Aseev2018Lifetime}  are not expected (see Fig.~\ref{figPhaseDiagramEnergyDifference}). 
Although in the general case this quantity does not provide direct information about the topology of the phase, in a system with no other sub-gap states than MBSs, this energy difference is given by the finite-size level-spacing in the trivial phase and is very small. In contrast to that, in the topological phases, it determines the value of the topological mini-gap, which should be much larger than the level spacing (see, respectively, the dark red regions of Fig.~\ref{figPhaseDiagramEnergyDifference}).  As expected, the critical value of the continuum model $ J_C $ (yellow line Fig.~\ref{figPhaseDiagramEnergyDifference}), agrees well for small values of the rotation rate $\Phi$ with the almost zero value of $E_2-E_1$ and with $J^<_C$ (orange line Fig.~\ref{figPhaseDiagramEnergyDifference}). The upper critical value of the lattice model, $J^{>}_C$, confines the topological phase space in the regime of strong $J$. For example, in the case of an antiferromagnetic alignment the topological phase is absent \cite{Kobialka2021Majorana,Diaz2021Majorana}.  We also note that the interval spanned by the values of $J^{<}_C(\Phi)$ and $J^{>}_C(\Phi)$ is maximal in case of ferromagnetic ordering. However, the system with ferromagnetic configuration stays trivial since the bulk gap does not reopen, for more details see  Ref. \cite{Perge2013Proposal}.\\

We now define the different non-uniform rotation profiles that we will use in the remainder of the manuscript. In general, these non-uniform rotation rates will involve a transition between two different rotation rates in a short and long section of the atomic chain, mathematically described by the profile defined in Eq.~\eqref{eq:AngleOmega}. We refer to a section as {\it short} if its length is comparable or shorter than the smoothness parameter $\lambda$ (see above) and a section as {\it long} if its length is larger than $\lambda$. 
Since the phase diagram shown in Fig.~\ref{figPhaseDiagramEnergyDifference} verifies that the topological phase transition criterion $J^<_C$ is strongly dependent upon the rotation rate $\Phi_n$ between neighbouring magnetic moments, we can expect a rich variety of states arising due to these non-uniform profiles. Throughout we will consider only one transition region on the left end of the chain. However, most probably, both ends should be identical, such that there is also a trivial state on the right. We consider only one region such that we can demonstrate the difference between a uniform chain end and one with the transition region at the end.

The first non-uniform rotation rate profile we define is that of a \textit{smooth decay}, this is shown in Fig.~\ref{figAngleSmoothDecay}. In this profile, the magnetic moments rotate in the longer section of the chain with a rate $\Phi_R<\Phi_{k_F}$, as such this section of the chain obeys the topological phase transition criterion for exchange couplings larger than $J^<_C(\Phi_R)$. In addition, the magnetic moments in the short section, e.g. on the left side of the chain, rotate slower with a rate $\Phi_L<\Phi_R$ and, as a result, the topological phase transition criterion in this section is shifted to $J^{<}_C(\Phi_L)$ which is larger than $J^<_C(\Phi_R)$. Consequently, the entire system stays in the trivial regime for exchange couplings smaller than $J^{<}_C(\Phi_R)$ and any sub-gap state in this regime must have a trivial origin. An example of such a profile would be a ferromagnetic ordering of the magnetic moments close to the left end of the chain and a helical ordering on the right end. The consequences of this profile will be discussed in detail in Sec.~\ref{Sec:SmoothDecay}.\\

Another rotation rate profile is one that contains a \textit{domain wall} i.e. a change of the rotation direction, for example from clockwise to anticlockwise rotation. For instance, a rotation rate, which is negative in a short section on the left ($\Phi_L<0$), but which takes the value $\Phi_R>0$ with
\begin{align}
 \Phi_{k_F}>\Phi_R\geq |\Phi_L|\gg \frac{2\pi}{N_C} \label{eq:DomainWallRotationRates}
\end{align}
 in the longer right section of the atomic chain. The arguments from above apply also in this case and the topological phase transition criterion is shifted to  values of the exchange coupling larger than $J^{<}_C(\Phi_R)$ in the region where the sign of $\Phi_n$ changes, see Fig.~\ref{figAngleDomainWall}. The entire system therefore stays trivial below for $J<J^{<}_C(\Phi_R)$ and any sub-gap states  in this limit must be of trivial origin.  We refer to Sec.~\ref{Sec:DomainWall} for a detailed study of the sub-gap states resulting from such a profile. \\

Next, we consider a similar setup as in the case of the smooth decay, however, for this profile $\Phi_R<\Phi_L<\Phi_{k_F}$ (see Fig.~\ref{figAngleQuasiMBSFromBelowJT}). This choice has a significant impact on $J^{<}_C$, namely, the critical value of the exchange coupling for which the system obeys the topological phase transition criterion in this case is smaller for the left section with the fast rotating magnetic moments than for the long section of the chain with the slow rotating magnetic moments. Therefore, when $ J^{<}_C(\Phi_L)\leq J<J^{<}_C(\Phi_R)$ the short section on the left of the atomic chain nominally could enter the topological regime. However, these  MBSs will be strongly overlapping spatially and, thus, be hybridized. In Rashba nanowires zero-energy bound states arising due to a section of the system entering the topological regime have been termed as 
quasi-MBSs \cite{Prada2020From} and we therefore refer to this as the quasi-MBS profile. In fact, several other quasi-MBS profiles are possible. For example, if the rotation rate is set to $\Phi_L<\Phi_{k_F}$ in the short section on the left side of the chain and to  $\Phi_R>\Phi_{k_F}$ in the longer section on the right side, such that the rotation rate changes smoothly between these sections, then at some site $n=\tilde{n}$ of the chain $\Phi_{n=\tilde{n}}=\Phi_{k_F}$ (see Fig.~\ref{figAngleQuasiMBSCrossingkF}). As such, a subsystem, namely the  section where $ \Phi_n $ grows,  enters the topological regime for exchange couplings satisfying $J>\Delta$, $J<J^{<}_C(\Phi_L)$, and $J<J^{<}_C(\Phi_R)$.  We will investigate quasi-MBS profiles in Sec.~\ref{Sec:QuasiMBS}.\\

In summary the configurations, shown in Figs.~\ref{figAngleSmoothDecay} and \ref{figAngleDomainWall} only host trivial states since the exchange coupling is always smaller than the topological phase transition criterion $J^{<}_C(\Phi_R)$. In contrast, short sections of the systems shown in  Figs.~\ref{figAngleQuasiMBSFromBelowJT}-\ref{figAngleQuasiMBSCrossingkF} could obey the topological phase transition criterion locally. We will see that in such profiles a subsystem hosts hybridized (quasi-)MBSs. We note that topology is defined for bulk systems, so in general the topological phase transition criterion, $J^{<}_C$, is not meaningful for a single site or small section of a chain. Nevertheless, interpreting the behaviour of energies and wavefunctions  of sub-gap states in terms of the variation of the critical exchange coupling $J^{<}_C$, such that a section of the system enters the topological regime, agrees well with our numerical observations.

\section{Smooth Decay \label{Sec:SmoothDecay}}
This section deals with trivial zero-energy states in the one-dimensional helical spin chain model described in Sec.~\ref{Sec:ModelOneDimensionDiscretized} using the \textit{smooth decay} rotation rate profile, as described in Sec.~\ref{Sec:PhaseDiagram} (Fig.~\ref{figAngleSmoothDecay}). For this profile, in the long right section  of the chain the rotation rate $\Phi_n$ between adjacent magnetic moments is approximately constant such that $\Phi_n\approx\Phi_R$, however, the rotation rate $\Phi_n$ decreases to zero close to the left end of the chain (see blue line in Figs.~\ref{figProbDens_Aligned} and \ref{figProbDens_Embedded}).  \\
\subsection{One-dimensional model \label{Sec:1dModelSmoothDecay}}

\begin{figure}[t]
\subfloat{\label{figEnergiesAligned}\stackinset{l}{-0.00in}{t}{0.0in}{(a)}{\stackinset{l}{1.67in}{t}{0.0in}{(b)}{\stackinset{l}{-0.0in}{t}{0.95in}{(c)}{\stackinset{l}{1.67in}{t}{0.95in}{(d)}{\stackinset{l}{0.0in}{t}{1.86in}{(e)}{\stackinset{l}{1.67in}{t}{1.86in}{(f)}{\includegraphics[width=1\columnwidth]{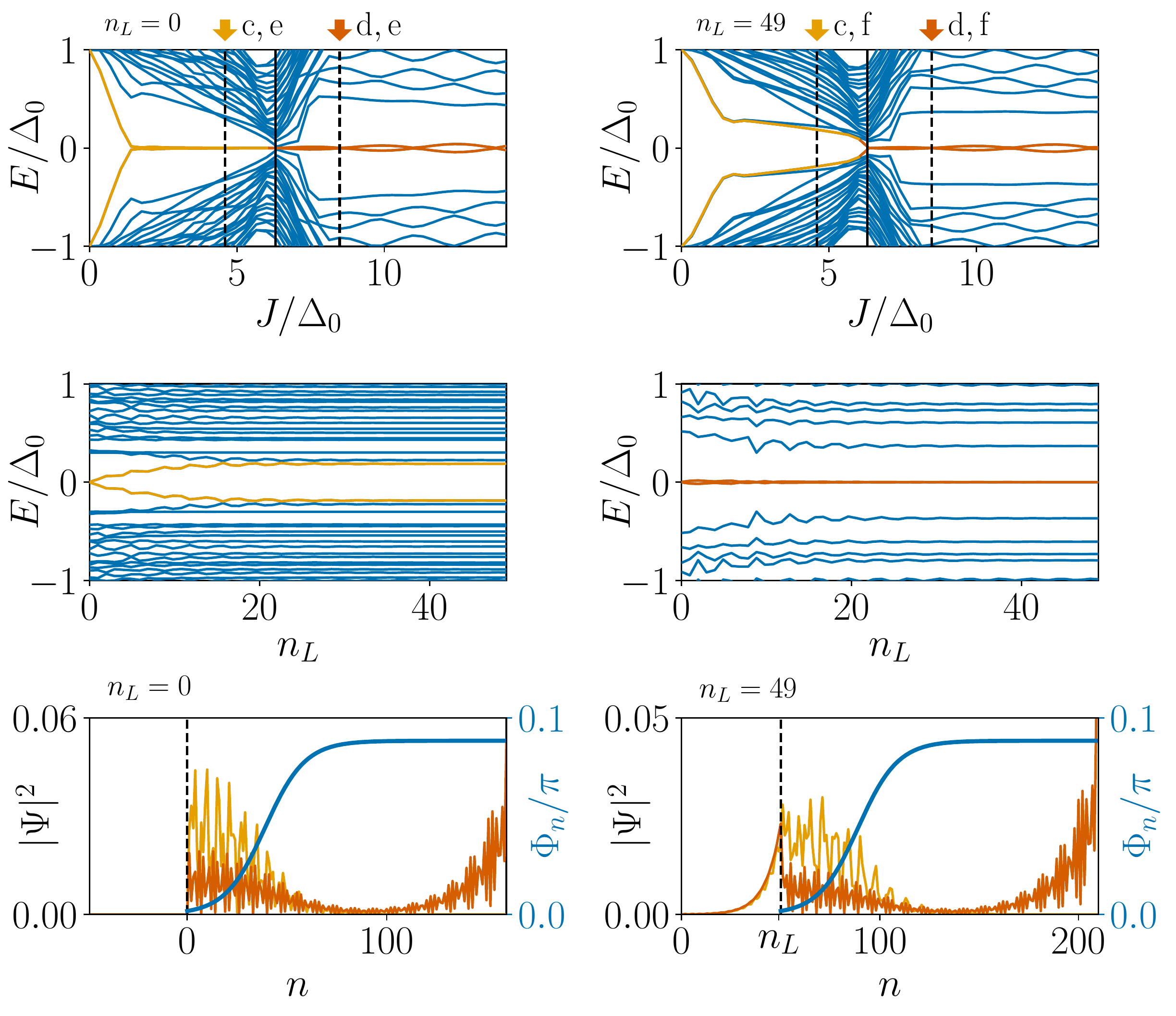}}}}}}}
}
\subfloat{\label{figEnergiesEmbedded}}
\subfloat{\label{figEnergiesTrivialTransition}}
\subfloat{\label{figEnergiesTopologicalTransition}}
\subfloat{\label{figProbDens_Aligned}}
\subfloat{\label{figProbDens_Embedded}}
 \caption{ \textit{Energy spectrum and probability density of the lowest energy state of a chain with a \underline{\text{smooth decay}} rotation profile on a $\underline{\text{one}}$-dimensional superconductor.}  (a,b) Energy spectrum as a function of the exchange coupling $J$ for  (a) the aligned and (b) embedded systems. The black vertical solid line  indicates the value $J=J^{<}_C(\Phi_R)$ at which the topological phase transition occurs. States in the completely trivial regime [$J<J^{<}_C(\Phi_R)$] are marked yellow, while the lowest sub-gap states in the topological regime [$J>J^{<}_C(\Phi_R)$] are marked orange. Higher energy states are shown in blue.  The helical spin chain hosts trivial zero-energy states only  in the aligned system but these states are continuously pushed away from zero energy as the number of sites without magnetic moments  $n_L$ on the left of the atomic chain grows. 
 (c) [(d)] Evolution of the  lowest trivial (yellow) [topological  (orange)] sub-gap state  as well as of the higher states as a function of the distance of the chain end from the system boundary at the exchange coupling strength  indicated by the yellow [orange] arrows in panels (a) [(b)]. 
 While the trivial sub-gap states split away from zero energy, see panel (c), the MBSs are stable and do not substantially change in energy, see panel (d). 
 (e) [(f)] The probability densities of the trivial (yellow) and the topological (orange) lowest-energy states calculated at the exchange couplings indicated by the yellow (orange) arrows in panels (a) and (b) in case of an aligned [embedded] system. The blue line  represents the rotation rate $\Phi_n$ between adjacent magnetic moments and the black dashed line indicates the end of the chain. 
Moreover, the probability densities show a clear difference between the MBSs and the trivial sub-gap states, and the latter are approximately confined in the  left section.
Parameters: $N_C=160$, $t\approx 10$ meV, $\mu=7$ meV, $\Delta=1$ meV, $\Phi_L=0$, $2\Phi_R=0.1768\pi$, $\lambda=20$, $n_0=40$, $a=3 \, \mbox{\normalfont\AA}$, $n_R=0$. }
 \label{figPaper1DHelicalChainAlignedVsEmbedded1}
\end{figure}

The energy spectrum of the one-dimensional lattice model with a smooth decay rotation profile is shown in Fig.~\ref{figEnergiesAligned} (aligned) and Fig.~\ref{figEnergiesEmbedded} (embedded) as a function of exchange coupling $J$. In both cases we find that the bulk gap closes and reopens for $J\approx J^{<}_C(\Phi_R)$, indicated by a solid black vertical line. In the regime $J<J^{<}_C(\Phi_R)$ we find that sub-gap states appear. These sub-gap states must be entirely trivial in nature, since the topological phase transition criterion is not met for any section of the chain. 

When the chain ends coincide with the boundaries of the superconductor (aligned system), the lowest energy sub-gap state is localized and pinned to zero energy over a range of exchange couplings (see Fig.~\ref{figEnergiesAligned}). Such a zero-energy bound state is reminiscent of an MBS, but here is entirely trivial. Since the exchange coupling $J<J_C(\Phi)$ does not satisfy the topological criterion in Eq.~\eqref{eq:BothTPTs} for any $\Phi$, the existence of such a zero-energy state is surprising and only arises here due to the non-uniformity of the rotation profile.
However, when the chain is embedded into the superconductor, such that the length of the superconductor far exceeds the length of the chain, we find that the zero-energy pinning is lifted and the sub-gap states split away from zero energy (see Fig.~\ref{figEnergiesEmbedded}). 

To further analyze the behavior of the low energy sub-gap states, we examine the transition from an aligned ($n_{L}=0$) to an embedded ($n_{L} a\gg  \xi$) system by adding lattice sites to the left of the atomic chain and by calculating the energy spectrum as a function of the distance $n_{L}$ between the  chain end and boundary of the superconductor. Here, $\xi=\hbar v_F/\Delta_0$ denotes the superconducting coherence length with $v_F$ being the Fermi velocity.  The resulting energy spectra for an exchange coupling smaller than $J^{<}_C(\Phi_R)$, indicated by the yellow arrow, and for an exchange coupling larger than $J^{<}_C(\Phi_R)$, indicated by the orange arrow, are shown in Figs.~\ref{figEnergiesTrivialTransition} and \ref{figEnergiesTopologicalTransition}, respectively. This transition from the aligned to the embedded system shows that the trivial zero-energy state is continuously pushed away from zero energy as superconductor  sites are added to the end of the chain. In addition, the next highest state decreases in energy such that the two lowest sub-gap states become approximately degenerate for a sufficiently long section without magnetic adatoms on the left of the chain (see Fig.~\ref{figEnergiesTrivialTransition}). In contrast, the zero-energy pinning of the MBS is essentially unaffected by an increase in $n_L$ (see Fig.~\ref{figEnergiesTopologicalTransition}). Analyzing the probability density $|\Psi (n)|^2$, see Figs.~\ref{figProbDens_Aligned} and \ref{figProbDens_Embedded}, reveals that the trivial state and the left MBS tend to leak into the section without magnetic adatoms in the case of an embedded system. 
More specifically, the wavefunction $\Psi  (n)$ decays exponentially into the section without magnetic moments with  $|\Psi|^2 \sim e^{-2na/\xi}$ \cite{Klinovaja2012Composite,Vernek2014Subtle}.\\
 
Hence already in one dimension we observe that, in spite of the mapping of Sec.~\ref{Sec:Model}, the leakage of the lowest energy states of an atomic chain into the surrounding superconductor results in an increase in energy of that state. 
 In general, mechanisms that suppress the leakage of the wavefunction into the section without magnetic adatoms can restore the zero-energy pinning of this lowest state.  For instance, in one dimension, a scalar impurity at the end of the chain reduces the leakage and a strong impurity results in a lowest energy state that is again pinned close to zero energy (not shown). 

Reducing the smoothness of the transition profile $ \Phi_{n}(\lambda) $ via the parameter $ \lambda $ [see Eq.~\eqref{eq:AngleOmega}], lifts the zero-energy pinning of the lowest sub-gap state in the topologically trivial regime. In contrast, the energies of the MBSs are unaffected by the smootheness of the profile $ \Phi_{n}(\lambda) $. The exact  form of the function $ \Phi_{n}(\lambda)$ is actually not crucial for the existence of the zero-energy sub-gap states  as long as $\Phi_{n}(\lambda)$ changes sufficiently smoothly between $\Phi_L$ and $\Phi_R$.
 In particular, we find that $\lambda$ needs to be larger than the rotation period of the magnetic moments on the right end of the atomic chain to obtain a sub-gap state with energy pinned to zero. Furthermore, the position $ n_0 $ in Eq.~\eqref{eq:AngleOmega} of the smooth step should be placed at least twice the length of $\lambda$ from the site of  first magnetic adatom and the larger the value of $n_0$  the more sub-gap states enter the spectrum. \\

Trivial sub-gap states appear at finite values of exchange coupling. By analyzing our numerical simulations we find that their energy is only pinned to zero when the condition 
\begin{align}
\Delta_0<J<J^<_C(\Phi_R)  
\label{eq:conditionForSubGapStates}
\end{align}
holds. We can understand this behaviour in the picture outlined in Ref.~\citenum{kells2012Near} for semiconductor nanowires: when the exchange coupling $J$ is larger than the superconducting gap, the superconducting correlation has an effective  $p$-wave nature and below the critical exchange coupling $J^<_C(\Phi_R)$ two $p$-wave channels are present.  The smooth change of the rotation rate, which maps to a smooth variation of potential and induced SOI in the effective model, does not allow a coupling between the two channels and supports the formation  of trivial zero-energy states. In contrast, an abrupt change in the spatial profiles couples the two channels and the sub-gap states are pushed away from zero energy. On the other hand, for $J>J^<_C (\Phi_R)$, the sub-gap states evolve into MBSs \cite{kells2012Near}.

 If the rotation rate is set to $\Phi_R= \Phi_{k_F}$ in the longer section of the chain, as predicted for an ordering mediated by the RKKY interaction \cite{Rudermann1954Indirect,Kasuya1956Theory,Yosida1957Magnetic,Choy2011Majorana, Martin2012Majorana,Perge2013Proposal,Pientka2013Topological, Pientka2014Unconventional, Reis2014SelfOrganized,  Poeyhoenen2014Majorana, Klinovaja2013TopologicalSuper}, and if the magnetic moments deviate from the $k_F$ ordering on the very left end of the chain, as $\Phi_L<  \Phi_{k_F}$, then almost the entire system, except the  section on the left, satisfies the topological phase transition criterion when  $J_C(k_r=k_F)\geq\Delta_0 $, see Eq.~\eqref{eq:PhaseTransitionContinous}. Consequently, the interval  of exchange coupling strengths for which trivial sub-gap states might appear shrinks to zero [see  Eq.~\eqref{eq:conditionForSubGapStates}]. Therefore, if the magnetic moments in the long right  section of the chain form a spiral with period $\pi/k_F$, as  RKKY interaction suggests, then no sub-gap states are present even in the case of an aligned system. In contrast, if the spin ordering deviates from $k_r=k_F$ in a long section of the chain, see for example Refs.~\cite{Schecter2015SpinLattice}, then the formation of zero-energy states of trivial nature is possible for the aligned system.

\subsection{Two-dimensional model \label{Sec:2dModelSmoothDecay}}

\begin{figure}[t]
\subfloat{\label{figEnergiesAligned2d}\stackinset{l}{-0.00in}{t}{0.in}{(a)}{\stackinset{l}{1.67in}{t}{0.in}{(b)}{\stackinset{l}{-0.0in}{t}{0.88in}{(c)}{\stackinset{l}{1.67in}{t}{0.88in}{(d)}{\stackinset{l}{0.0in}{t}{1.8in}{(e)}{\stackinset{l}{1.75in}{t}{1.8in}{(f)}{\includegraphics[width=1\columnwidth]{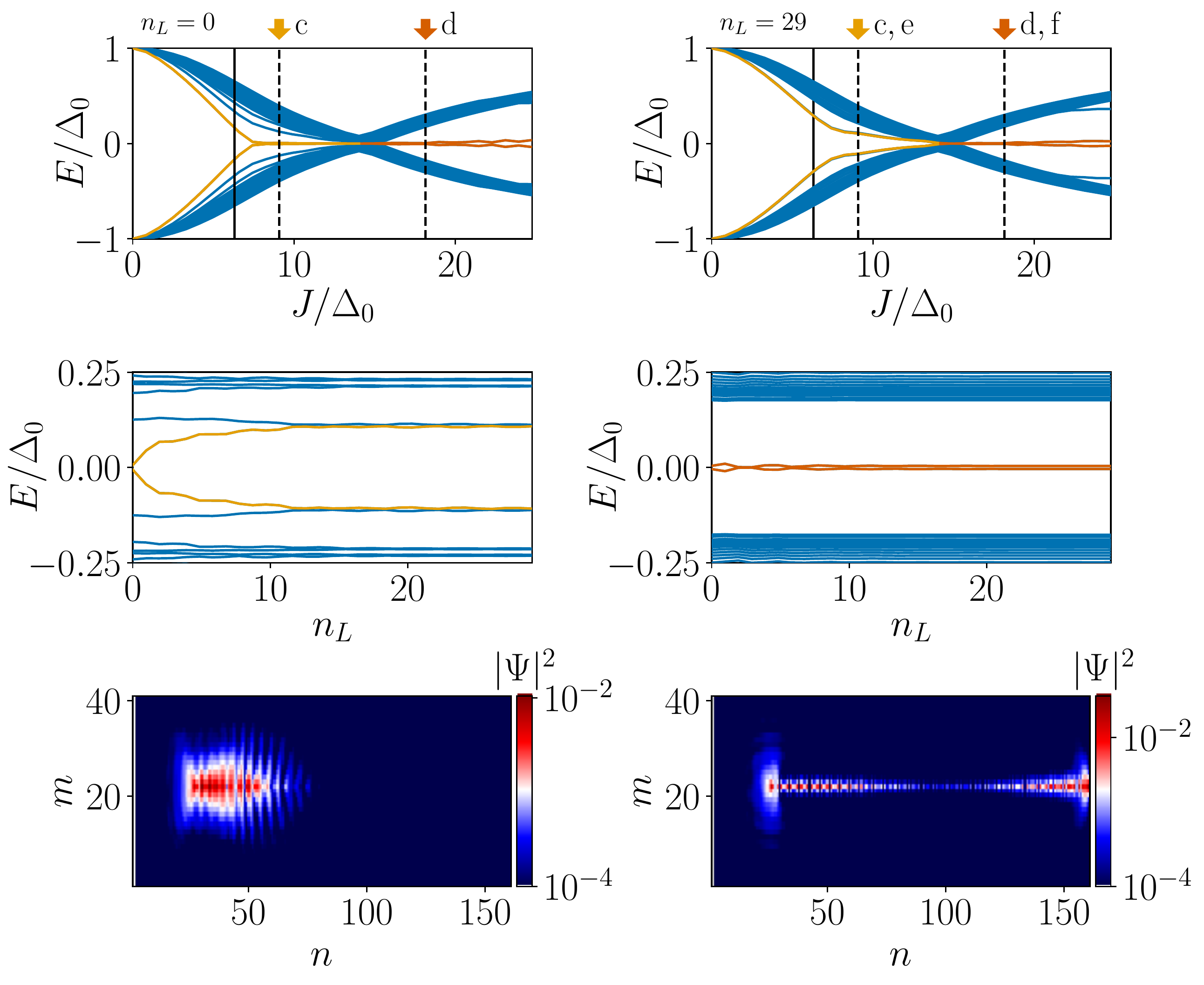}}}}}}}
}
\subfloat{\label{figEnergiesEmbedded2d}}
\subfloat{\label{figEnergiesTrivialTransition2d}}
\subfloat{\label{figEnergiesTopologicalTransition2d}}
\subfloat{\label{figProbDens_trivial2d}}
\subfloat{\label{figProbDens_topological2d}}
 \caption{\textit{Energy spectrum and probability density of the lowest energy state of a chain with a \underline{\text{smooth decay}} rotation profile on a $\underline{\text{two}}$-dimensional superconductor.} (a,b) Energy spectrum as a function of the exchange coupling $J$ for  (a) the aligned and (b) embedded systems. The black vertical solid line  indicates the value $J=J^{<}_C(\Phi_R)$ in the analogous one-dimensional system, this does not match the actual bulk gap closing and reopening observed in two-dimensional setup.
  (a) The lowest sub-gap state has almost zero energy when the chain is aligned (b) but the state has finite energy when the chain is embedded. (c) [(d)] Evolution of the  lowest trivial (yellow) [topological  (orange)] sub-gap state  as well as of the higher states as a function of the distance of the chain end from the system boundary at the exchange coupling strength  indicated by the yellow [orange] arrows in panel (a).  (c) The  lowest trivial state is continuously pushed away from zero energy by adding more non-magnetic sites to the left of the end of the chain. (d) In contrast the energy of the MBSs is essentially unaffected by the number of non-magnetic sites at the end of the chain. (e) [f] The probability densities of the trivial (yellow) and the topological (orange) low-energy states calculated at the exchange couplings indicated by the yellow (orange) arrows in panels (a) and (b) in case of an aligned [embedded] system.
 The wavefunction of the lowest energy trivial sub-gap state is mostly localized at the section of the atomic chain with the slow-rotating magnetic moments but it leaks also slightly (strongly) in the $x$ ($y$) direction parallel (perpendicular) to the chain into the section with no magnetic adatoms. In contrast, the probability density of the lowest state in the topological phase reveals two peaks at the ends of the chain due to the left and right MBS.
The parameters are the same as in Fig.~\ref{figPaper1DHelicalChainAlignedVsEmbedded1}. We choose $N_y=41$ and place the chain on the line $m_0=21$ along the $y$-direction.}
 \label{HelicalChainDirectVsIndirectDependenceInTwoDimensionsSmoothDecay}
\end{figure}

We now consider a helical magnetic atomic chain placed on top of a two-dimensional superconductor, as presented in Sec.~\ref{Sec:2dAnd3dModel}. 
We choose the same smooth profile for the rotation rate between adjacent magnetic moments as in the purely one-dimensional system discussed above. Many of our findings are the same as for the one-dimensional system. For instance, a smooth decay profile leads to sub-gap states with an energy close to zero over some range of exchange coupling strengths in the trivial phase in the case of an aligned system (see Fig.~\ref{figEnergiesAligned2d}). Extending the superconductor in the direction parallel to the chain (increasing $n_L$) lifts the zero-energy pinning of the trivial state, see Fig.~\ref{figEnergiesTrivialTransition2d}, in the same manner as in the embedded one-dimensional system. Moreover, the two lowest-energy states become almost degenerate in the trivial phase.  Again, the energy of the MBSs is nearly independent of $n_L$ (see Fig.~\ref{figEnergiesTopologicalTransition2d}) as in the one-dimensional system.

One significant difference to the analogous one-dimensional system is that the exchange coupling strength for which the bulk gap closes and reopens is larger than in the one-dimensional system \cite{Bjoernson2019Majorana}, see the black vertical solid line in Figs.~\ref{figEnergiesAligned2d} and \ref{figEnergiesEmbedded2d}. This discrepancy can be understood due to the additional leakage of the wavefunction  into the direction perpendicular to the chain (see wavefunction in Fig.~\ref{figProbDens_topological2d}). Due to this leakage, low-energy states have a weaker overlap with the sites where a finite exchange coupling is present, therefore the effective exchange coupling strength decreases and the topological phase transition is shifted to higher exchange coupling strengths $J$.  This effect is also reminiscent of the renormalization of the effective $g$-factor in Rashba or TI nanowires due to metallization caused by the coupling to a thin superconducting shell \cite{Cole2015Effects,Reeg2017Transport,Reeg2018Metallization,Legg2022Metallization}.
 We note that a leakage of the wavefunction of the trivial sub-gap states  in the perpendicular direction, see Fig.~\ref{figProbDens_trivial2d},  in the aligned system does not significantly change the zero-energy pinning, while the leakage in the direction  parallel of the chain  in the embedded system does affect the energy of the sub-gap states.

In the most experimentally realistic setup of an embedded chain on the superconductor, we do not find any zero-energy trivial sub-gap states. We also note that, unlike in the one-dimensional system, even a single-site scalar impurity at the end of the chain is now not sufficient  to restore the zero-energy pinning since, in two-dimensional systems, the confined state can bypass the impurity.

\subsection{Three-dimensional model \label{Sec:3dModelSmoothDecay}}

 \begin{figure}[!t]
\subfloat{\label{fig3D_SmoothDecay_EnergySpec}\stackinset{l}{0.in}{t}{0in}{(a)}{\stackinset{l}{0.in}{t}{1.25in}{(b)}{\stackinset{l}{0.2in}{t}{-1.1in}{\includegraphics[width=0.48\columnwidth]{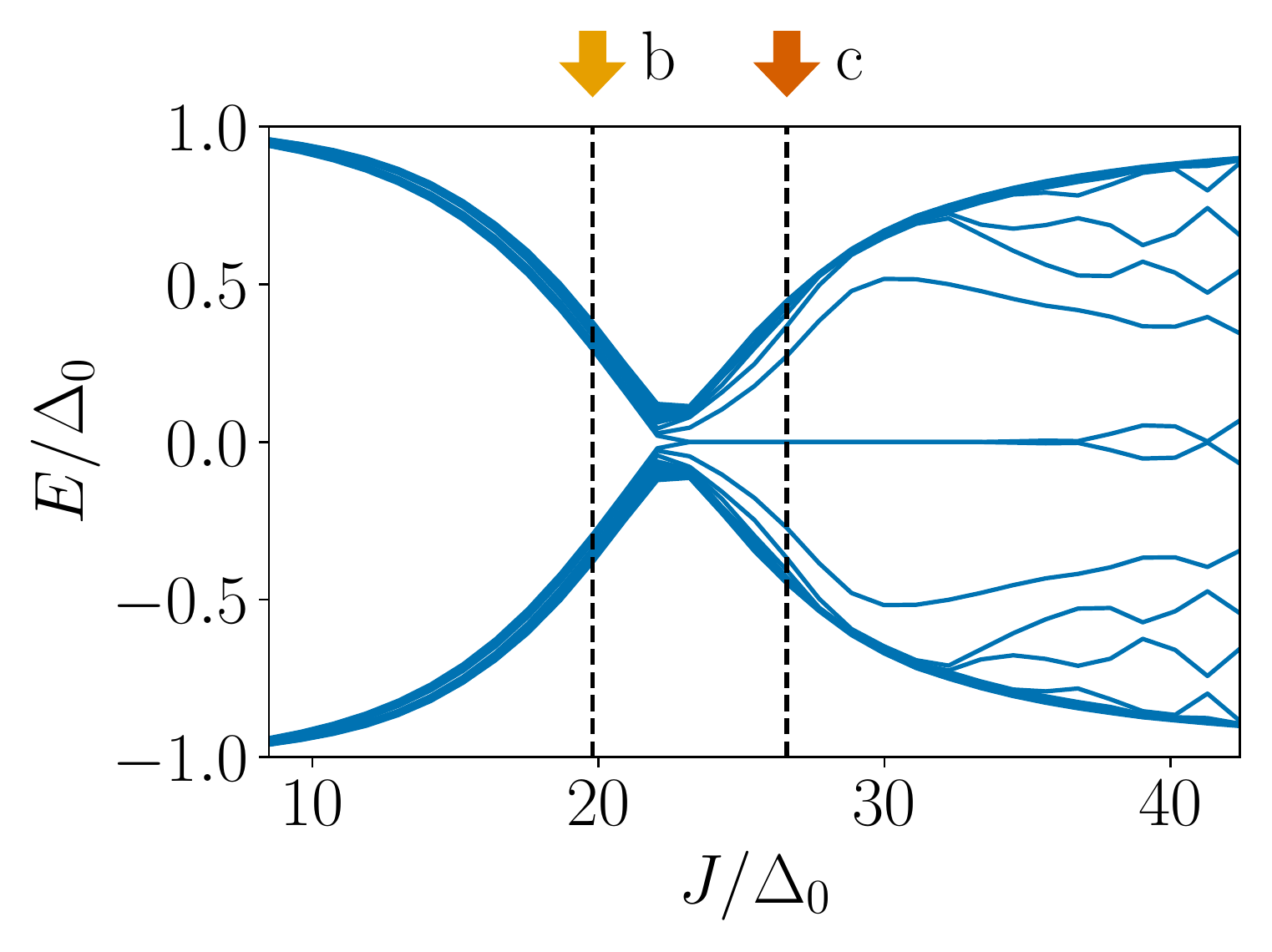}}{\includegraphics[width=1\columnwidth]{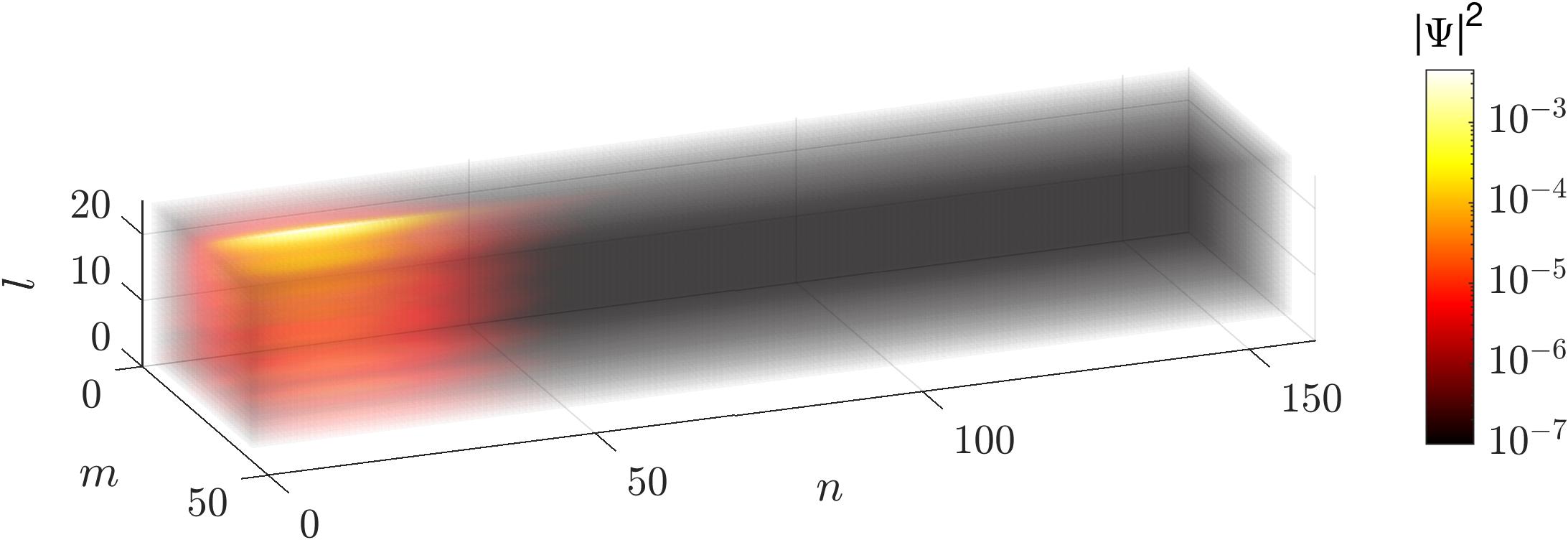}}}}
}\\
\vspace{-0.145in} 
\subfloat{\label{fig3dProbDensTrivialSmoothDecay}}
\subfloat{\label{fig3dProbDensTopoSmoothDecay}\stackinset{l}{0.0in}{t}{0.2in}{(c)}{\includegraphics[width=1\columnwidth]{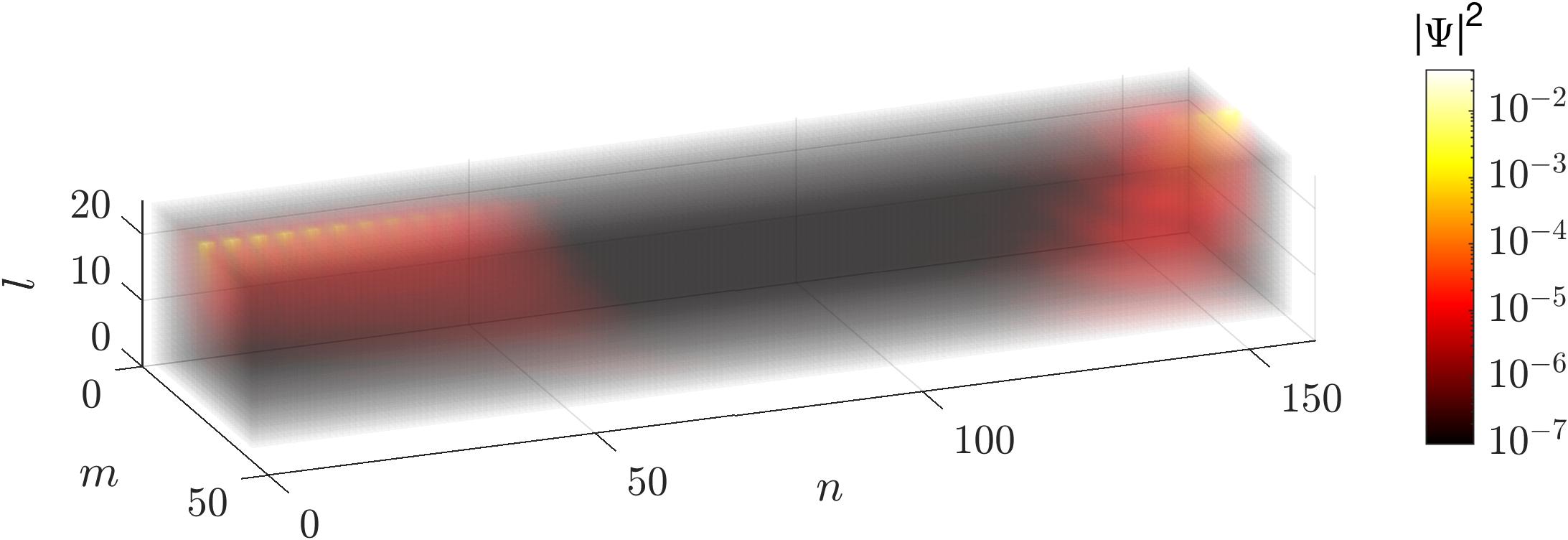}}
}
 \caption{{\it  Energy spectrum and probability density of the lowest energy state of a chain with a \underline{\text{smooth decay}} rotation profile on a $\underline{\text{three}}$-dimensional superconductor.} (a) Energy spectrum as a function of the exchange coupling $J$ for the aligned systems. The system undergoes a topological phase transition indicated by the gap closing and reopening and the appearance of zero-energy MBSs. We do not find trivial zero-energy sub-gap states in this three-dimensional system. (b,c) Probability density of the lowest energy state in the trivial and  the  topological regime, respectively. The lowest state in the trivial regime is mostly localized in the section of the smooth change of the rotation rate.  The probability densities are plotted for the exchange couplings strengths indicated by the yellow and orange arrow  in panel (a).  The probability density in (b) shows that the lowest state is localized in the region where the rotation rate changes, as expected from the one- and two-dimensional case, even though the state is energetically hardly separated from the bulk states. The MBSs are well separated and localized at the opposite ends of the chain. Again, their wavefuction is mostly located on the sites not covered by magnetic impurities, allowing to effectively diminish their localization length \cite{Zyuzin2013Correlations,Peng2015Strong}. The parameters are the same as in Fig.~\ref{figPaper1DHelicalChainAlignedVsEmbedded1}. In addition we choose $N_y=41$, $m_0=21$, $N_z=25$, and $l_0=1$. 
 }
 \label{fig3dSmoothDecay}
\end{figure}

Finally, we consider an atomic chain placed on the surface of a three-dimensional superconductor, utilising the model described in Sec.~\ref{Sec:2dAnd3dModel}. This model is closest to realistic experimental set-ups, in which the adatoms are deposited on top of a bulk three-dimensional superconductor \cite{Menzel2012Information,Howon2018Toward,Schneider2021Topological}. 

As was found in the two-dimensional model, also for the three-dimensional system there are no zero-energy states when the chain is embedded on the surface of the superconductor. Unlike the two-dimensional system, however, in the three-dimensional system we also do not find near zero-energy states when the chain end is  aligned with the boundary of the superconductor.  In particular, Fig.~\ref{fig3D_SmoothDecay_EnergySpec} shows the energy spectrum of an aligned system as a function of the exchange coupling. The gap closes and reopens for an exchange coupling strength $J_C$ that is larger than observed in both the one-dimensional and two-dimensional systems. For coupling strengths below the value at which the gap closes we do not observe any zero-energy sub-gap states, even in the aligned system.  We also note that the curvature of the bulk gap closing lines differs from the curvature of the same process in  the two-dimensional superconductor (see Figs.~\ref{figEnergiesAligned2d} and \ref{figEnergiesEmbedded2d}). After the gap closing, as expected, we find MBSs appear at the end of the chain.

In the trivial regime, i.e. before the closing of the bulk gap, of the aligned chain the probability density of the lowest state is bound to the section in which $\Phi_n$ smoothly changes as we observed in one-  and two dimensional systems (see Fig.~\ref{fig3dProbDensTrivialSmoothDecay}), however the energy of this lowest state is essentially equal to that of the bulk gap. Both the lowest energy state in the trivial regime and the MBS decay exponentially into the bulk superconductor (see Figs.~\ref{fig3dProbDensTrivialSmoothDecay} and \ref{fig3dProbDensTopoSmoothDecay}).  

We note that much of the behavior of the aligned three-dimensional system can be understood from the fact that the chain is placed on the surface of the three-dimensional superconductor and therefore states can scatter from the surface into the bulk of the superconductor (along the $z$-direction). To investigate the importance of a boundary of the system perpendicular to the chain, we construct an analogous system in two dimensions, where the chain has the same length as the  superconductor in $x$ direction ($N_C=N_x$) and in which the chain is  placed along the boundary of the superconductor (see Appendix~\ref{App:2dChainOnEdge}).  In this scenario we also do not find trivial zero-energy sub-gap states, similar to the three-dimensional system. In contrast, when the chain is placed in the bulk of the superconductor rather than on the surface, then trivial sub-gap states with energies smaller than the bulk gap do appear (see Appendix~\ref{App:ChainIn3dBulk}). Both of these results highlight the importance of scattering from the boundary in pushing trivial zero-energy states to higher energies. 

We would like to emphasize that numerical restrictions limit us to relatively small systems in three dimensions and we cannot therefore  perform extensive numerical investigations in such systems. Nonetheless, the disappearance of trivial zero-energy and sub-gap states in our three dimensional set-up, even for the aligned chain, indicates that the formation of trivial zero-energy sub-gap states is strongly suppressed when the chain is placed on top of a three-dimensional superconductor.  Especially since zero-energy states do form in an aligned two-dimensional system of comparable width and length. As the most realistic experimental setup is an embedded  chain on the surface of a three-dimensional superconductor  we can conclude that there is a low prevalence of zero-energy sub-gap states due to a smooth decay rotation profile.

\section{Domain wall \label{Sec:DomainWall}}

In this section, we will investigate the sub-gap states forming due to domain-wall rotation profiles. The results from the previous section show that the formation of trivial zero-energy states is unlikely when the length of the superconductor ($x$ direction) exceeds the length of the chain and there is a smooth decay in rotation rate. However, sub-gap states due to a domain wall profile, i.e. a smooth change between a clockwise and anticlockwise rotation of the magnetic moments within the chain, can be expected to be largely independent of the relative position of the chain end and the superconductor boundary. Although it should be noted that, for this profile, low energy subgap states can be expected to form close to the domain wall rather than at ends of the chain, as is expected for MBSs. 

\subsection{One-dimensional model \label{Sec:1dModelDomainWall}}

\begin{figure}[t]
\subfloat{\label{figEnergiesAlignedDomainWall}\stackinset{l}{-0.00in}{t}{-0.in}{(a)}{\stackinset{l}{1.67in}{t}{-0.in}{(b)}{\stackinset{l}{-0.0in}{t}{0.95in}{(c)}{\stackinset{l}{1.67in}{t}{0.95in}{(d)}{\stackinset{l}{0.0in}{t}{1.82in}{(e)}{\stackinset{l}{1.67in}{t}{1.82in}{(f)}{\includegraphics[width=1\columnwidth]{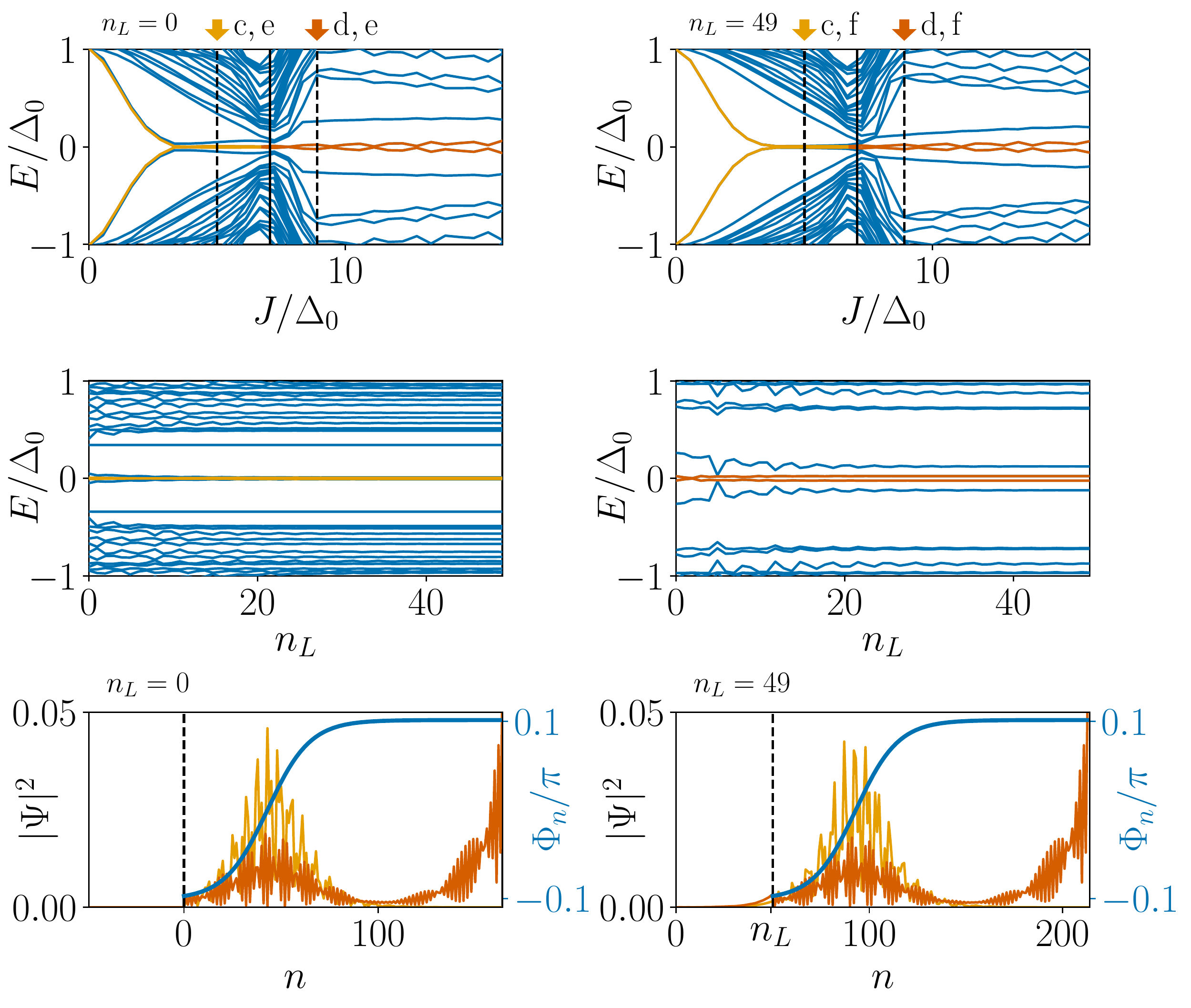}}}}}}}

}
\subfloat{\label{figEnergiesEmbeddedDomainWall}}
\subfloat{\label{figEnergiesTrivialTransitionDomainWall}}
\subfloat{\label{figEnergiesTopologicalTransitionDomainWall}}
\subfloat{\label{figProbDens_AlignedDomainWall}}
\subfloat{\label{figProbDens_EmbeddedDomainWall}}
 \caption{ \textit{Energy spectrum and probability density of the lowest energy state of a chain with a \underline{\text{domain wall}} rotation profile on a $\underline{\text{one}}$-dimensional superconductor.}  The panels are arranged in the same manner as  in Fig.~\ref{figPaper1DHelicalChainAlignedVsEmbedded1}.  The energies of the trivial sub-gap states can be pinned to zero in case of (a) the aligned system and, in contrast to the previous system, also in case of (b) the embedded system. The domain wall allows the formation of a total of four MBSs, which hybridize and form multiple sub-gap states in the topological regime. (c,d) The trivial state (yellow) does not split away from zero when non-magnetic sites are added to the left of the helical spin chain and therefore behaves similarly to the lowest MBS (orange). (e,f) The wavefunction of the trivial  state (and the left MBS) is mainly localized in the region where the direction of the rotation changes ($\Phi_n=0$). The wavefunction profiles are similar in both cases. In particular, the wavefunction of the trivial state (yellow) is approximately zero in the section without magnetic adatoms. Parameters: $N_C=164$, $t\approx 10$ meV, $\mu=8$ meV, $\Delta=1$ meV, $2\Phi_L=-0.2026\pi$, $2\Phi_R=0.2026\pi$, $\lambda=22$, $n_0=44$, $a=3 \, \mbox{\normalfont\AA}$, $n_R=0$.}
 \label{figPaper1DHelicalChainAlignedVsEmbeddedDomainWall1}
\end{figure}

First, utilising the one-dimensional model outlined in Sec.~\ref{Sec:ModelOneDimensionDiscretized} we find that a chain with a domain wall profile does support trivial states with almost zero energy in both the case of an aligned (see Fig.~\ref{figEnergiesAlignedDomainWall}) and also an embedded chain (see Fig.~\ref{figEnergiesEmbeddedDomainWall}), which is in contrast to the system with the smooth decay profile, for which we did not observe trivial zero-energy states in the embedded case.
In particular, within a domain wall profile, the rotation rate of magnetic moments is set by Eqs.~(\ref{eq:AngleOmega}) and (\ref{eq:DomainWallRotationRates})  such that the rotation rate of the magnetic moments smoothly interpolates  between $\Phi_L\ll - \frac{2 \pi}{N_C}$ on the left side and $\Phi_R\gg \frac{2 \pi}{N_C}$ on the right side of the chain. Here, we chose for simplicity $\Phi_L=-\Phi_R$, which means that the direction of the rotation changes along the atomic chain (see blue lines in Figs.~\ref{figProbDens_AlignedDomainWall} and \ref{figProbDens_EmbeddedDomainWall}).   As such, since the critical exchange coupling at which the gap closes is shifted to larger values for slower rotating magnetic moments, the chain is completely trivial for an exchange coupling below $J^{<}_C(\Phi_R)$ (see also Fig.~\ref{figAngleDomainWall}). Within this trivial regime, we find  two sub-gap states -- as well as their particle-hole partners -- that have almost zero energy and the transition from the aligned to the embedded system does not substantially affect this zero-energy pinning (see Fig.~\ref{figEnergiesTrivialTransitionDomainWall}). Similar to the chain with a smooth decay, the two lowest sub-gap states of the embedded chain become almost degenerate in the case that the section without magnetic atoms to the left of the atomic chain is sufficiently long, such that the localization length of the states is much shorter than the length of the left section of the superconductor without any magnetic adatoms. 

Further analyzing the wavefunction of the states in this trivial regime $J<J^{<}_C(\Phi)$, we find the maximum of the probability density of the sub-gap states is localized at the position of the sign change of $\Phi_n$ and that the probability density is almost zero in the region where the rotation rate reaches its maximal positive or maximal negative value (see blue line Fig.~\ref{figProbDens_AlignedDomainWall}). Adding  more sites to the left of the chain does not substantially affect the energy spectrum since the weight of the wavefunction in the section without magnetic atoms is very small (see  Fig.~\ref{figProbDens_EmbeddedDomainWall}). In the topological regime, i.e. after the closing and reopening of the bulk gap, we find the system hosts four MBSs.  In particular, one MBS appears at each end of the chain and one MBS on each side of the domain wall. Depending on the length of the domain wall transition the MBSs closest to the domain wall can hybridize and form a finite-energy sub-gap state \cite{Ojanen2013Topological,Klinovaja2015Fermionic, Rossi2020Confinement, Ronetti2020Magnetically}.

\subsection{Two-dimensional model \label{Sec:2dModelDomainWall}}
\begin{figure}[t]
\subfloat{\label{figEnergiesAligned2dDomainWall}\stackinset{l}{-0.00in}{t}{-0.in}{(a)}{\stackinset{l}{1.67in}{t}{-0.in}{(b)}{\stackinset{l}{-0.0in}{t}{0.88in}{(c)}{\stackinset{l}{1.67in}{t}{0.88in}{(d)}{\stackinset{l}{0.0in}{t}{1.8in}{(e)}{\stackinset{l}{1.72in}{t}{1.8in}{(f)}{\includegraphics[width=1\columnwidth]{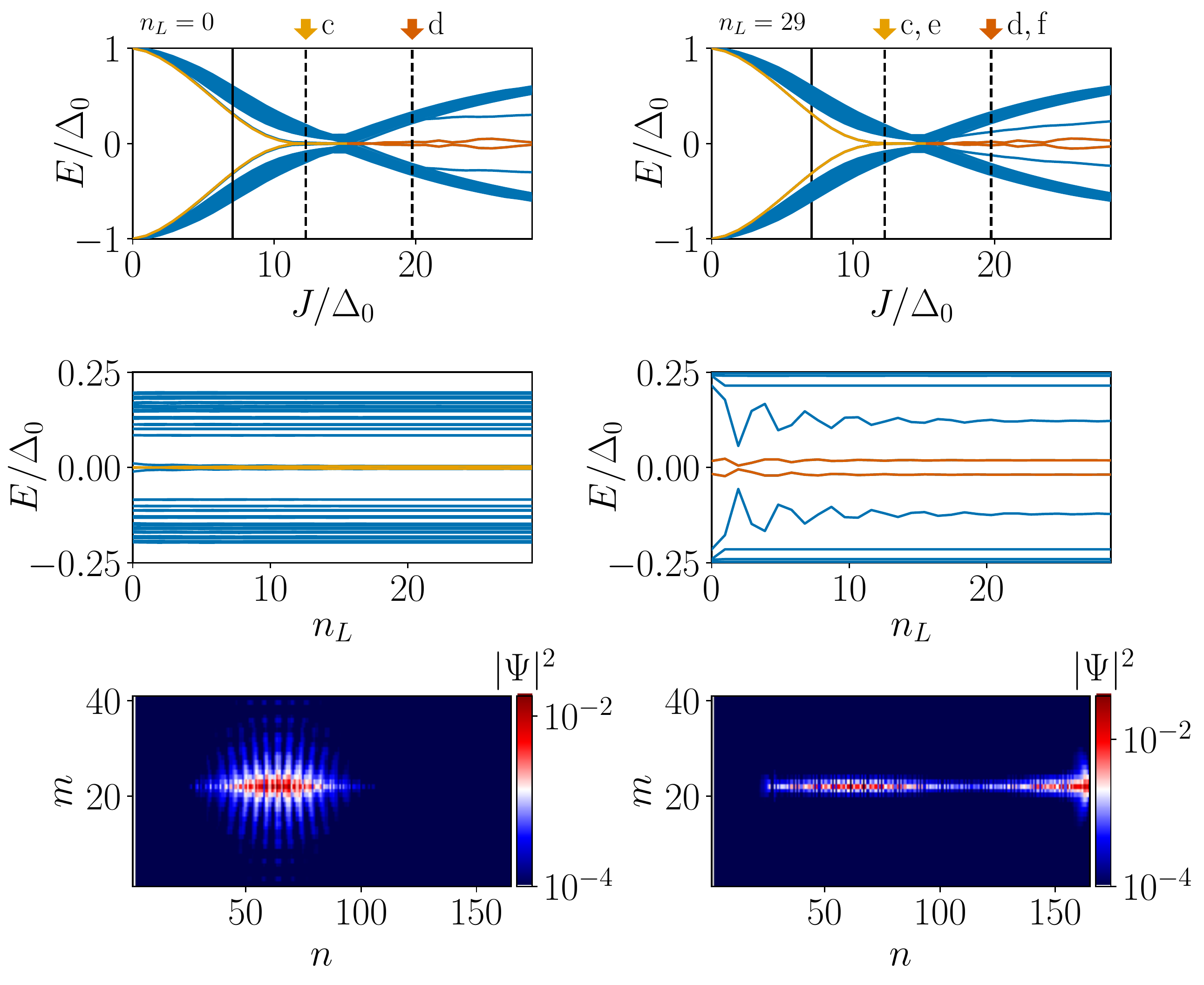}}}}}}}
}
\subfloat{\label{figEnergiesEmbedded2dDomainWall}}
\subfloat{\label{figEnergiesTrivialTransition2dDomainWall}}
\subfloat{\label{figEnergiesTopologicalTransition2dDomainWall}}
\subfloat{\label{figProbDens_trivial2dDomainWall}}
\subfloat{\label{figProbDens_topological2dDomainWall}}
 \caption{\textit{Energy spectrum and probability density of the lowest energy state of a chain with a \underline{\text{domain wall}} rotation profile on a $\underline{\text{two}}$-dimensional superconductor.}  The panels are arranged in the same manner as  in Fig.~\ref{HelicalChainDirectVsIndirectDependenceInTwoDimensionsSmoothDecay}.  The lowest sub-gap state has almost zero energy when (a) the chain is aligned and also when (b) the chain is embedded. The black solid vertical line indicates $J^{<}_C$ in an analogous one-dimensional system, the line does not match with the actual bulk gap closing and reopening in two dimension system.   The energy of the lowest state is neither affected  in (c) the trivial nor in (d) the topological regime  during the crossover from the aligned to the embedded system if the chain is long enough. (e) The wavefunction of the lowest energy trivial sub-gap state is mostly localized at the domain wall but also slightly leaks into the section with no magnetic adatoms perpendicular to the chain. In contrast in panel (f), the probability density of the lowest state in the topological phase reveals two additional  peaks at the ends of the chain due to the left and right MBS. Here, for our choice of parameters, the MBSs hybridize strongly.
     The parameters are the same as in Fig.~\ref{figPaper1DHelicalChainAlignedVsEmbeddedDomainWall1} and  we chose $N_y=41$ and $m_0=21$ to account for the two-dimensional system.}
 \label{HelicalChainDirectVsIndirectDependenceInTwoDimensionsSDomainWall}
\end{figure}

Next, we consider the helical chain with the domain wall profile that is placed on top of a two-dimensional superconductor. The physical properties of the sub-gap states are similar to those found in one-dimensional system, namely trivial states are pinned to zero-energy in both cases of an aligned and embedded chain (see Figs.~\ref{figEnergiesAligned2dDomainWall} and \ref{figEnergiesEmbedded2dDomainWall}). The lowest energy states in the trivial regime before the bulk gap closes and reopens (see Fig.~\ref{figEnergiesTrivialTransition2dDomainWall}), as well as in the topological regime after the closing and reopening of the bulk gap (see Fig.~\ref{figEnergiesTopologicalTransition2dDomainWall}), are unaffected by the value of  $n_L$. Furthermore, as found in the one-dimensional system, the probability density of the near zero-energy sub-gap states is localized to the region of the smooth change of the rotation rate (see Fig.~\ref{figProbDens_trivial2dDomainWall}). In the topological regime -- after the reopening of the bulk gap -- strongly overlapping MBSs appear,  their probability density is maximal on the right end and in the region of the smooth transition (see Fig.~\ref{figProbDens_topological2dDomainWall}).  We note that the wavefunction of the MBS  only weakly leaks into the $y$-direction perpendicular to the chain, such that the wavefunction is confined very close to the position of the chain.

\subsection{Three-dimensional model \label{Sec:3dModelDomainWall}}

 \begin{figure}[t]
\subfloat{\label{fig3D_DomainWall_EnergySpec}\stackinset{l}{0.in}{t}{0in}{(a)}{\stackinset{l}{0in}{t}{1.25in}{(b)}{\stackinset{l}{0.in}{t}{-1.1in}{\includegraphics[width=0.8\columnwidth]{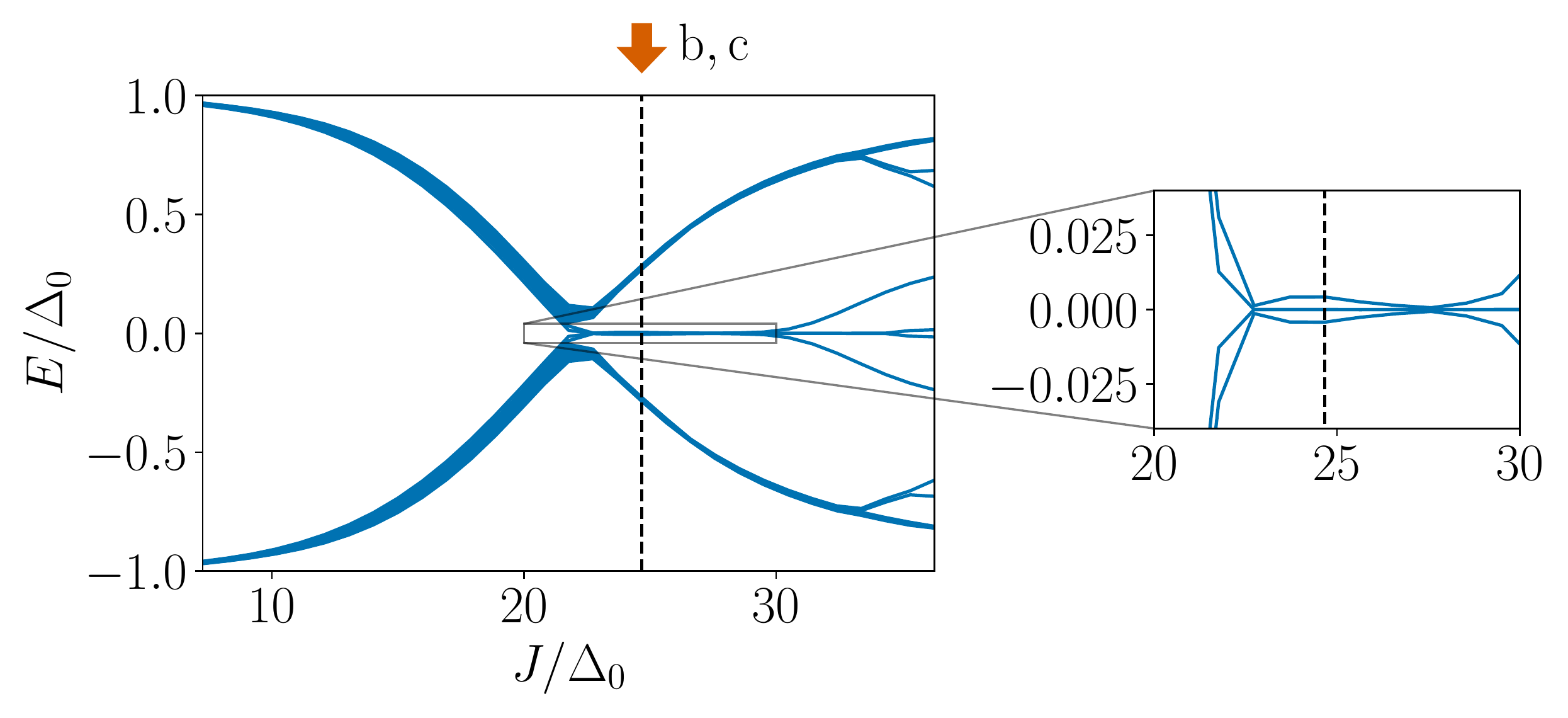}}{\includegraphics[width=1\columnwidth]{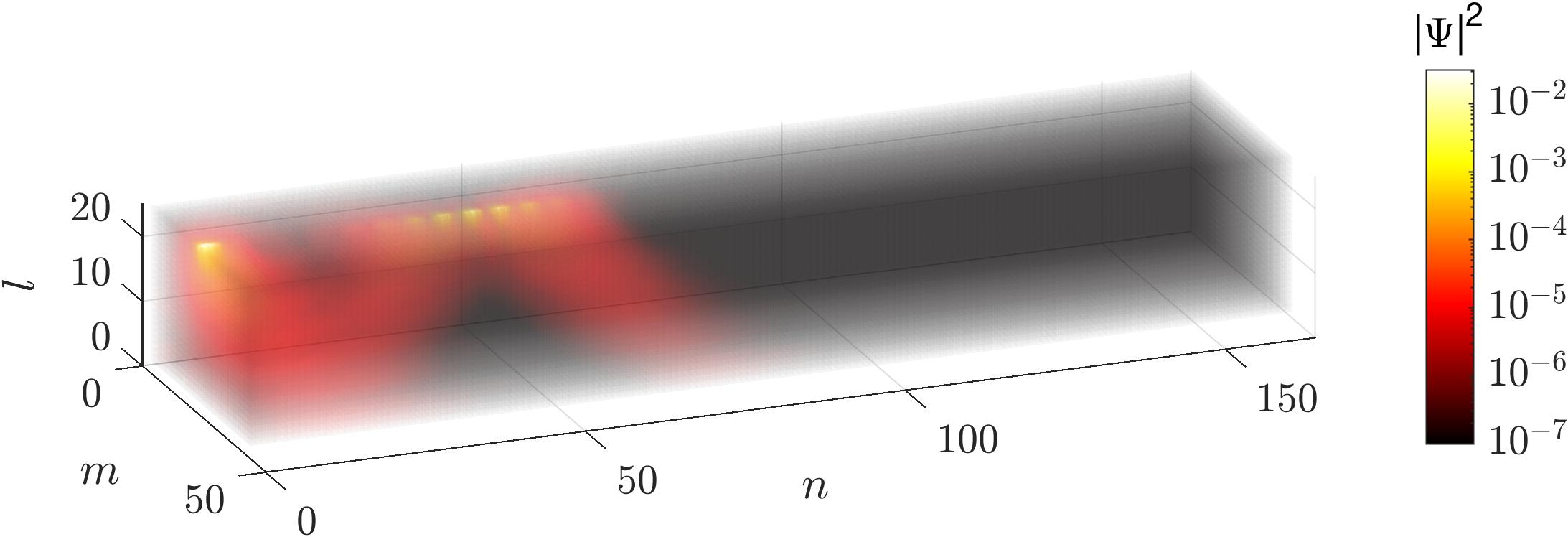}}}}
}\\
\subfloat{\label{fig3dProbDensTrivialDomainWall}}
\vspace{-0.145in}
\subfloat{\label{fig3dProbDensTopoDomainWall}\stackinset{l}{0.0in}{t}{0.2in}{(c)}{\includegraphics[width=1\columnwidth]{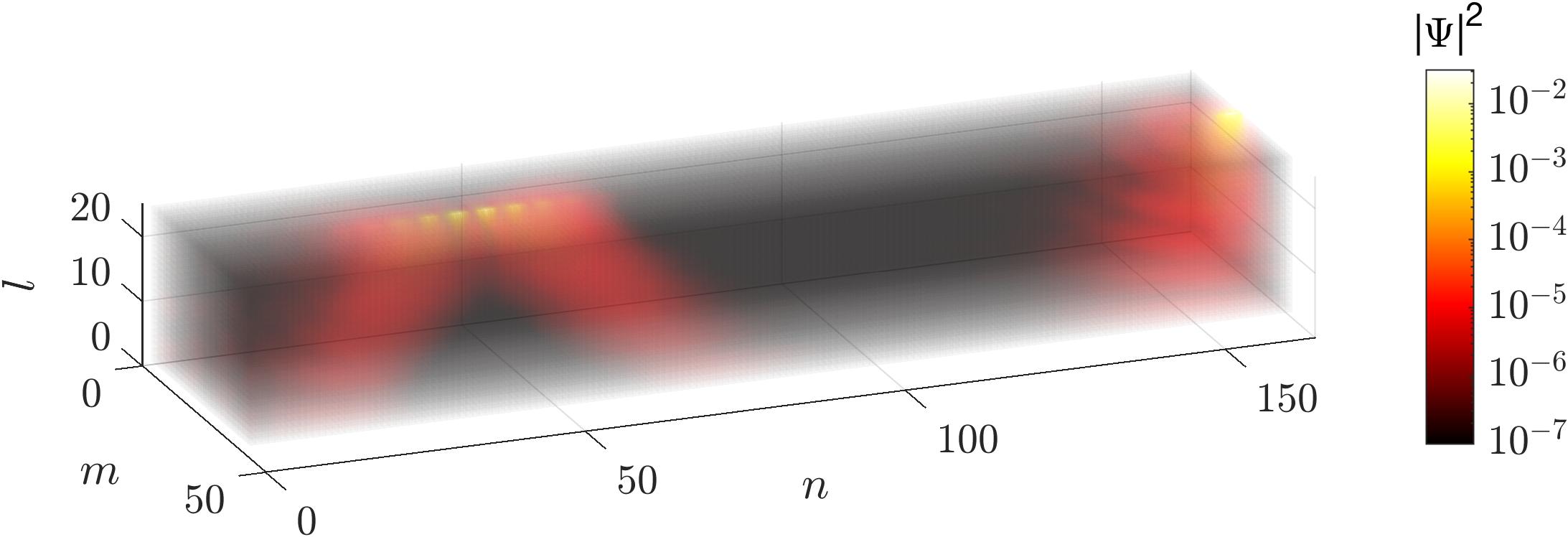}}
}
 \caption{\textit{Energy spectrum and probability density of the lowest energy state of a chain with a \underline{\text{domain wall}} rotation profile on a $\underline{\text{three}}$-dimensional superconductor.}  The panels are arranged in the same manner as  in Fig.~\ref{fig3dSmoothDecay}.  (a)  No sub-gap states are present before the topological phase transition, indicated by the gap closing and reopening. In the topological regime, MBSs appear. The system hosts four MBSs: two on each side of the domain wall. A zoom into the panel reveals the slightly different energies of the two MBS pairs close to zero due to finite overlap between MBS wavefunctions.  The probability density of (b) the second lowest  and (c) lowest  energy states in the  topological phase. The probability densities are plotted for the exchange couplings strength indicated by the orange arrow in panel (a).  The parameters are the same as in Fig.~\ref{figPaper1DHelicalChainAlignedVsEmbeddedDomainWall1} and in addition we chose $N_y=41$,  $m_0=21$, $N_z=25$, and $l_0=1$ to account for the $y$ and $z$ directions. }
 \label{fig3dDomainWall}
\end{figure}

In contrast to the one- and two-dimensional systems considered above, we find that a helical spin chain  with a domain wall spin profile that is placed on the surface of a three-dimensional superconductor does not exhibit trivial near zero-energy states. The energy spectrum is shown in Fig.~\ref{fig3D_DomainWall_EnergySpec}.  The MBSs appear after the reopening of the bulk gap but there are no sub-gap states in the trivial regime. This behaviour corresponds with a similar observation for the smooth decay profile with aligned boundaries (see Sec.~\ref{Sec:3dModelSmoothDecay}). In the three-dimensional system, the chain is placed on the surface of the superconductor, thus, boundary effects such as scattering from the surface also affect the energy of the lowest states.  Finally we note, as also observed in one- and two-dimensional systems, the probability density of the two lowest  states in the topological regime after the bulk gap has closed, reveals the presence of a total of four MBSs, with two that are localized at the ends of the chain and on either side of the domain wall. The MBSs close to the domain wall partially hybridize (see Figs.~\ref{fig3dProbDensTrivialDomainWall} and \ref{fig3dProbDensTopoDomainWall}).

\section{Quasi-MBS \label{Sec:QuasiMBS}}
In this section, we consider a chain in which only one  section of the chain nominally obeys the topological phase transition criterion, see the discussion in Sec.~\ref{Sec:PhaseDiagram} and Fig.~\ref{figAngleQuasiMBSFromBelowJT}. In particular, we consider a chain in which the magnetic moments rotate with the rate $\Phi_R$ in the long right section of the chain but in the left short section the rotation rate between neighbouring magnetic moments increases smoothly up to $\Phi_L$,  with $\Phi_{k_F}>\Phi_L>\Phi_R$. The zero-energy sub-gap states that result from a system only partially obeying the topological phase transition criterion in a certain range of exchange couplings (see Sec. \ref{Sec:QuasiMBS1D} for the specific values of $J$ in the one-dimensional system) have been termed \textit{quasi-MBSs} \cite{Prada2020From}.

\subsection{One-dimensional model\label{Sec:QuasiMBS1D}}

\begin{figure}[t]
\subfloat{\label{figEnergiesAlignedQuasiMBS}\stackinset{l}{-0.00in}{t}{-0.in}{(a)}{\stackinset{l}{1.67in}{t}{-0.in}{(b)}{\stackinset{l}{-0.0in}{t}{0.95in}{(c)}{\stackinset{l}{1.67in}{t}{0.95in}{(d)}{\stackinset{l}{0.0in}{t}{1.86in}{(e)}{\stackinset{l}{1.67in}{t}{1.86in}{(f)}{\includegraphics[width=1\columnwidth]{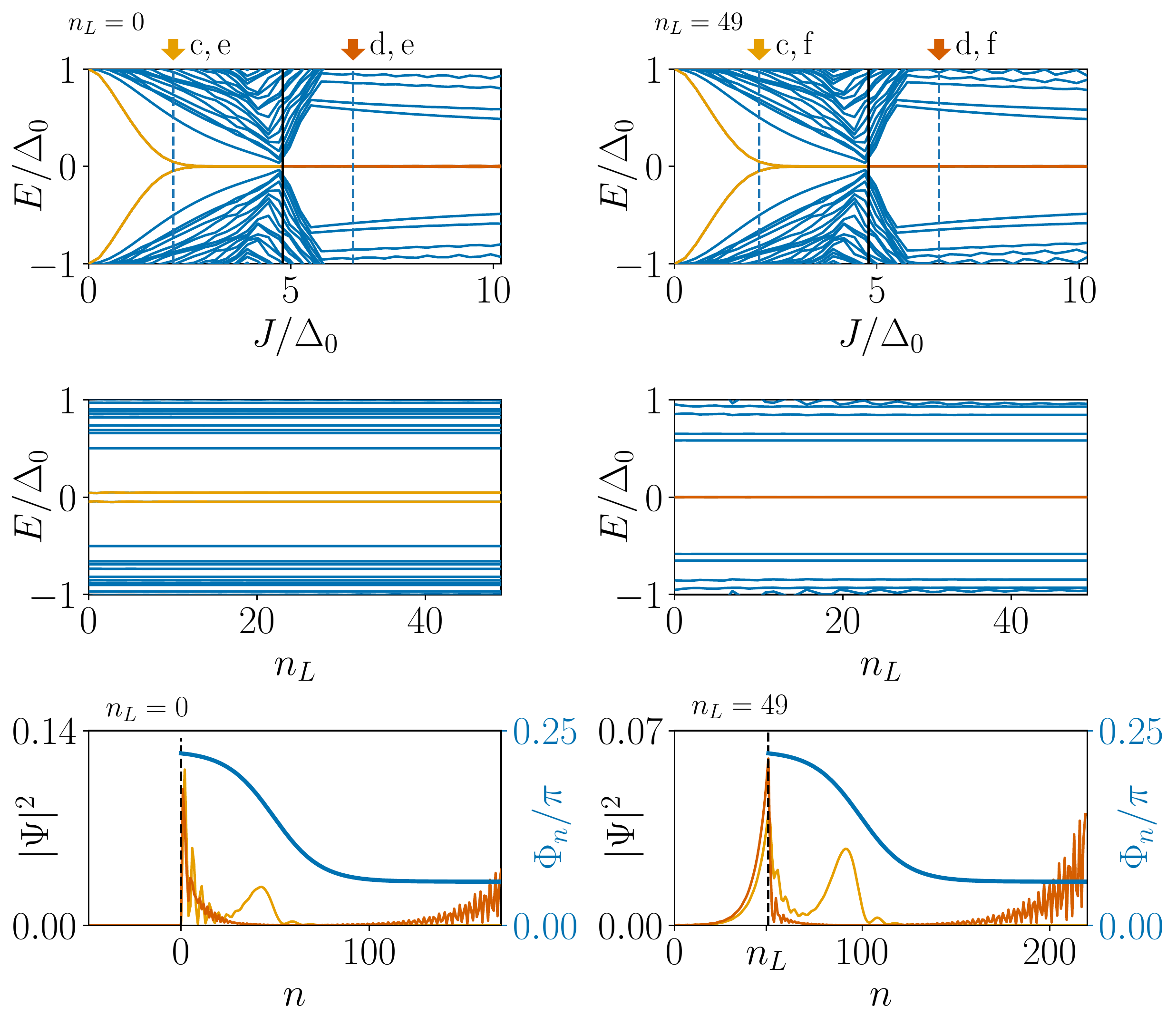}}}}}}}

}
\subfloat{\label{figEnergiesEmbeddedQuasiMBS}}
\subfloat{\label{figEnergiesTrivialTransitionQuasiMBS}}
\subfloat{\label{figEnergiesTopologicalTransitionQuasiMBS}}
\subfloat{\label{figProbDens_AlignedQuasiMBS}}
\subfloat{\label{figProbDens_EmbeddedQuasiMBS}}
 \caption{ \textit{Energy spectrum and probability density of the lowest energy state of a chain with a \underline{\text{quasi-MBS}} rotation profile on a $\underline{\text{one}}$-dimensional superconductor.}   The panels are arranged in the same manner as  in Fig.~\ref{figPaper1DHelicalChainAlignedVsEmbedded1}.  The system hosts the so-called quasi MBSs  (yellow) for  (a)  the aligned and (b) the embedded system. (c) [(d)] The energy of the quasi-MBS [the MBS] is nearly independent of the number of sites which are placed to the left of the chain, though the probability densities shown in panels (e) and (f) of the quasi-MBSs (yellow) and the MBSs (orange) are substantially shifted into the section without magnetic adatoms. The probability density reveals clearly that the quasi-MBS originates from hybridizing MBSs which form in the section of faster rotation. Parameters: $N_C=170$, $t\approx 10$ meV, $\mu=5$ meV, $\Delta=1$ meV, $2\Phi_L=0.4484\pi$, $2\Phi_R=0.1121\pi$, $\lambda=25$, $n_0=50$, $a=3 \, \mbox{\normalfont\AA}$, $n_R=0$.
  }
 \label{Paper1DHelicalChainAlignedVsEmbeddedQuasiMBS1}
\end{figure}

First, considering a one-dimensional system, we find the bulk gap closing and reopening appears at $J^{<}_C(\Phi_R)$. The shorter left section with the faster rotating magnetic moments, however, obeys the topological phase transition criterion for $J^{<}_C(\Phi_L)<J<J^{<}_C(\Phi_R)$ (see Fig.~\ref{figAngleQuasiMBSFromBelowJT}). 
 As a result, the system contains a sub-gap state for exchange couplings smaller than $J^{<}_C(\Phi_R)$ which results from the presence of two hybridizing MBSs. The energy of this state is well pinned to zero over a range of exchange coupling strengths in both the aligned (see Fig.~\ref{figEnergiesAlignedQuasiMBS}) and  in the embedded setup (see Fig.~\ref{figEnergiesEmbeddedQuasiMBS}). This effect is shown clearly in Fig.~\ref{figEnergiesTopologicalTransitionQuasiMBS} where we observe that the energy of the lowest sub-gap states below and above $J^{<}_C(\Phi_R)$ remain close to zero as an increasing number of sites $n_L$ without magnetic adatoms are added to the superconductor on the left of the system.  In contrast to the previous profiles we do not find a degeneracy of the two lowest energy states in the embedded system.

The spatial profile of wavefunctions of these sub-gap states reveals the MBS character of the quasi-MBSs. In particular, in the regime $J< J^{<}_C(\Phi_R)$, the probability density of the lowest state (yellow) has two separated peaks at the ends of the section that obeys the topological phase transition criterion (see Figs.~\ref{figProbDens_AlignedQuasiMBS} and \ref{figProbDens_EmbeddedQuasiMBS}). 

Increasing the exchange coupling to values of $J\approx J^{<}_C(\Phi_R)$ allows one to bring the entire chain into the topological regime. We find the quasi-MBSs transform into MBSs that are localised at the ends of the chain (see orange line in Figs.~\ref{figProbDens_AlignedQuasiMBS} and \ref{figProbDens_EmbeddedQuasiMBS}).  Other rotation rate profiles, which lead to a local reduction of the critical exchange coupling $J^{<}_C$, as suggested in Fig.~\ref{figAngleQuasiMBSCrossingkF}, support quasi-MBSs as well and lead to similar results as discussed in this section (not shown). We emphasize  that the mechanism for zero-energy states in partially topological chains crucially differs from the one in fully trivial chains, see Secs.~\ref{Sec:SmoothDecay}  and \ref{Sec:DomainWall}. However, similar to the domain wall case,  near zero-energy states appear in the partially topological chain independent of the boundary conditions for a purely one-dimensional system.

\subsection{Two-dimensional model \label{Sec:2dModelQuasiMBSs}}

\begin{figure}[t]
\subfloat{\label{figEnergiesAligned2dQuasiMBS}\stackinset{l}{-0.00in}{t}{-0.in}{(a)}{\stackinset{l}{1.67in}{t}{-0.in}{(b)}{\stackinset{l}{-0.0in}{t}{0.88in}{(c)}{\stackinset{l}{1.67in}{t}{0.88in}{(d)}{\stackinset{l}{0.0in}{t}{1.8in}{(e)}{\stackinset{l}{1.72in}{t}{1.8in}{(f)}{\includegraphics[width=1\columnwidth]{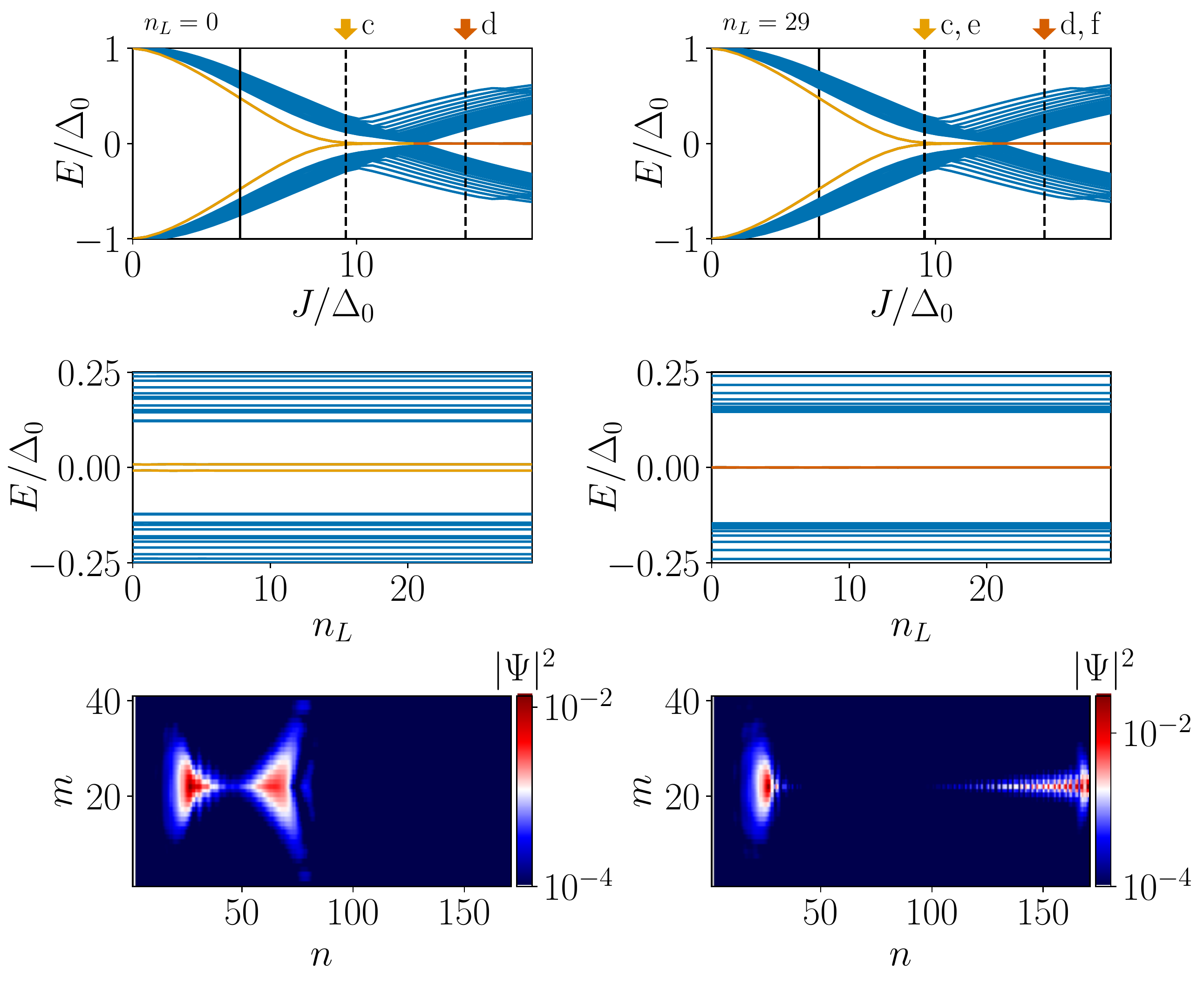}}}}}}}
}
\subfloat{\label{figEnergiesEmbedded2dQuasiMBS}}
\subfloat{\label{figEnergiesTrivialTransition2dQuasiMBS}}
\subfloat{\label{figEnergiesTopologicalTransition2dQuasiMBS}}
\subfloat{\label{figProbDens_trivial2dQuasiMBS}}
\subfloat{\label{figProbDens_topological2dQuasiMBS}}
 \caption{\textit{Energy spectrum and probability density of the lowest energy state of a chain with a \underline{\text{quasi-MBS}} rotation profile on a $\underline{\text{two}}$-dimensional superconductor.}  The panels are arranged in the same manner as  in Fig.~\ref{HelicalChainDirectVsIndirectDependenceInTwoDimensionsSmoothDecay}.  The main results are similar to ones found for the one-dimensional model: the quasi-MBS  energy is pinned close to zero in the case of (a) the aligned and (b) the embedded system. A calculation of the energy spectrum as a function of the length $n_L$ of the left section without magnetic adatoms  in (c) below  and (d) above the bulk gap closing and reopening shows that the energies are basically not affected, in contrast, to the systems discussed in Secs.~\ref{Sec:SmoothDecay} and \ref{Sec:DomainWall}. The probability density for exchange couplings (e) smaller and (f) larger than the coupling corresponding to the bulk gap closing has a MBS character: The chain can partially enter the topological phase due to the increasing rotation rate of the magnetic moments in the left section of the chain. Consequently, MBSs appear at the ends of this short topological section and move to the ends of the chain with increasing exchange coupling. 
The parameters are the same as in Fig.~\ref{Paper1DHelicalChainAlignedVsEmbeddedQuasiMBS1} and in addition we chose $N_y=41$ and $m_0=21$.}
 \label{figHelicalChainDirectVsIndirectDependenceInTwoDimensionsSQuasiMBSs}
\end{figure}

The behavior of a two-dimensional system, consisting of a one-dimensional partially topological chain placed on top of a two-dimensional superconductor, agrees well with the properties of the simple one-dimensional model, presented in the previous subsection. In particular, the energy of the sub-gap  states does not depend on $n_L $ (see Figs.~\ref{figEnergiesTrivialTransition2dQuasiMBS} and \ref{figEnergiesTopologicalTransition2dQuasiMBS}) and the sub-gap states are pinned close to zero-energy over some range of the exchange coupling strength before the gap closes and reopens (see Figs.~\ref{figEnergiesAligned2dQuasiMBS} and \ref{figEnergiesEmbedded2dQuasiMBS}). A subsystem of the chain enters the topological regime for exchange couplings smaller than the value at which the bulk gap closing and reopening appears, with quasi-MBSs emerging at the ends of this subsystem (see Fig.~\ref{figProbDens_trivial2dQuasiMBS}). The leakage of the wavefunction in parallel or perpendicular direction to neighbouring sites of the chain does not affect the zero-energy pinning. For sufficiently large exchange couplings the gap closes and reopens and  the right MBS is pushed to the end of the chain (see Fig.~\ref{figProbDens_topological2dQuasiMBS}).

\subsection{Three-dimensional model \label{Sec:3dModelQuasiMBSs}}

 \begin{figure}[t]

 \subfloat{\label{fig3dEnergySpecQuasiMBS}\stackinset{l}{0.in}{t}{0in}{(a)}{\stackinset{l}{0.in}{t}{1.25in}{(b)}{\stackinset{l}{0.2in}{t}{-1.1in}{\includegraphics[width=0.48\columnwidth]{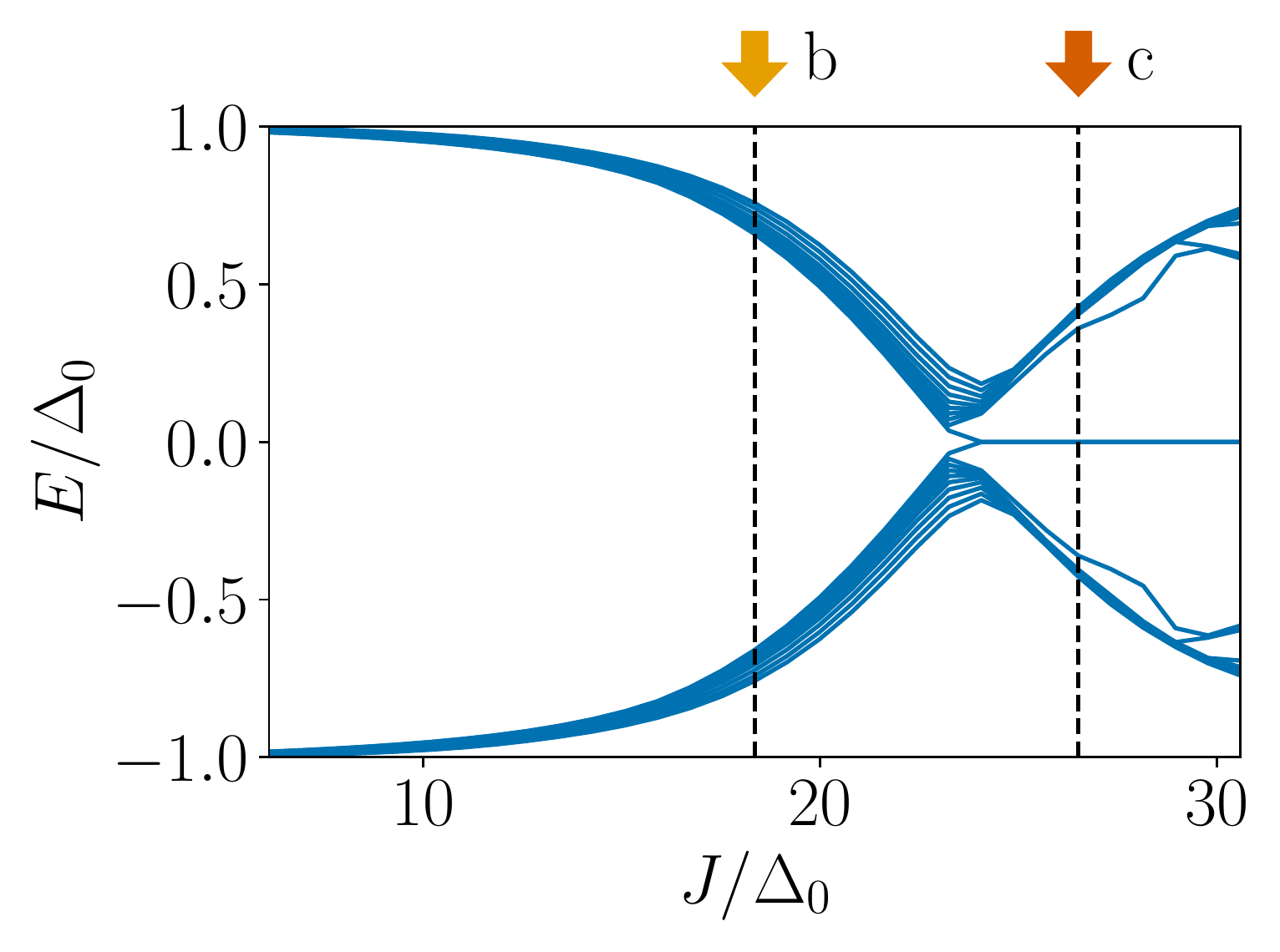}}{\includegraphics[width=1\columnwidth]{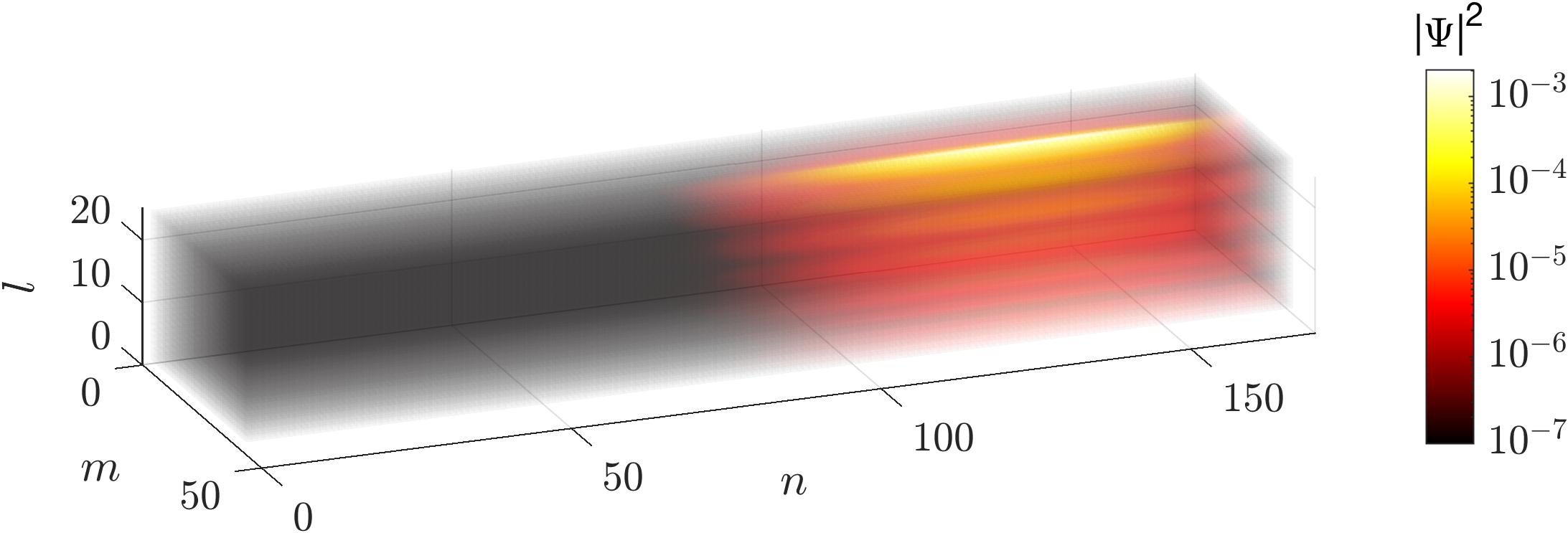}}}}
}\\
\subfloat{\label{figProbDens3d_QuasiMBS_Triv_2}}
\vspace{-0.145in}
\subfloat{\label{fig3dProbDensTopoQuasiMBS}\stackinset{l}{0.0in}{t}{0.2in}{(c)}{\includegraphics[width=1\columnwidth]{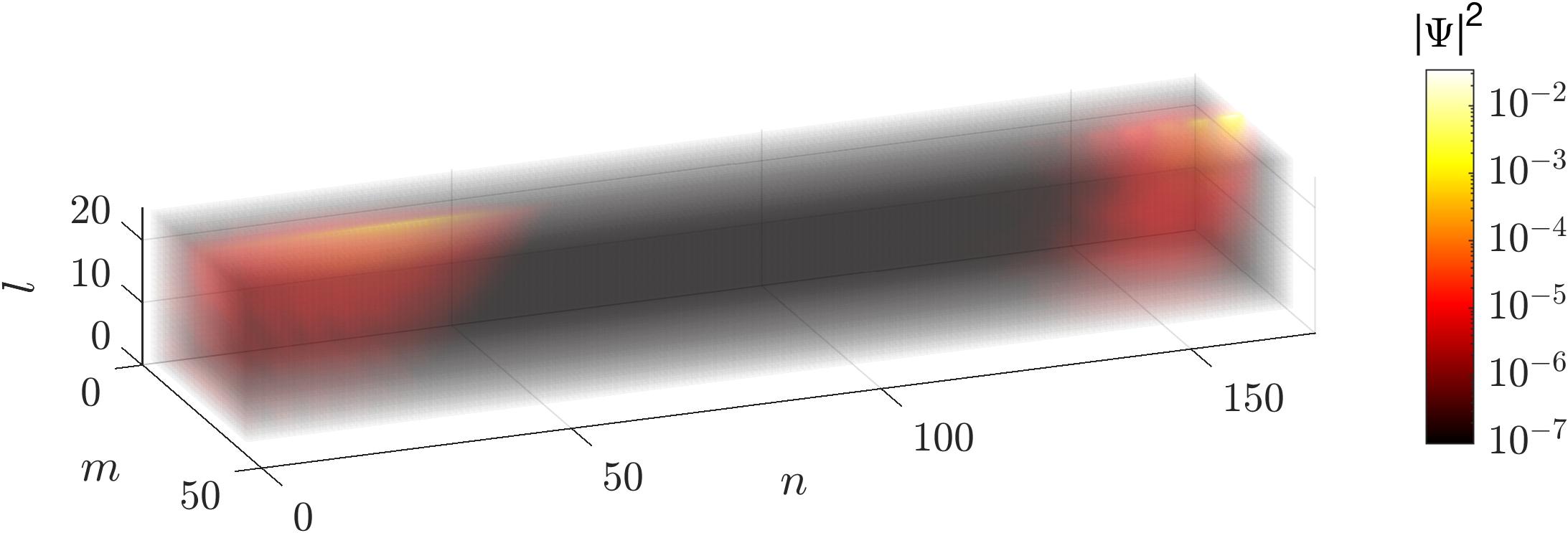}}
}
 \caption{\textit{Energy spectrum and probability density of the lowest energy state of a chain with a \underline{\text{quasi-MBS}} rotation profile on a $\underline{\text{three}}$-dimensional superconductor.}  The panels are arranged in the same manner as  in Fig.~\ref{fig3dSmoothDecay}.  (a)  The system undergoes a topological phase transition indicated by the gap closing and reopening and the appearance of zero-energy MBSs. This three-dimensional system does not host any trivial zero-energy sub-gap states.  Probability density of the lowest state energy in  (b) the trivial and  (c) the  topological phase, respectively. Unlike the one- and two-dimensional systems, the lowest state in the trivial state is not localized in the region of the smooth change of the rotation rate.  The probability densities are plotted for the exchange couplings strengths indicated by the yellow and orange arrow in panel (a).  The parameters are the same as in Fig.~\ref{Paper1DHelicalChainAlignedVsEmbeddedQuasiMBS1}  and in addition we choose $N_y=41$, $m_0=21$, $N_z=25$, $l_0=1$ }
 \label{fig3dQuasiMBSs}
\end{figure}

Finally, we examine a helical chain with a quasi-MBS rotation rate profile as in Fig.~\ref{figAngleQuasiMBSFromBelowJT} on top of a three-dimensional superconductor and use the same parameters for the chain as in the two-dimensional system. In contrast to the one- and two-dimensional systems we do not find near zero-energy states in this three-dimensional system, even though the parameters are the same. Indeed, the energy spectrum shows a clear bulk gap closing and reopening as a function of $J$ with zero-energy MBSs present only after this closing and reopening (see Fig.~\ref{fig3dEnergySpecQuasiMBS}).  The probability density of the lowest state in the trivial and the topological regime is shown in Figs.~\ref{figProbDens3d_QuasiMBS_Triv_2}  and \ref{fig3dProbDensTopoQuasiMBS}, respectively. The lowest-energy trivial state is not bound to the region in which $\Phi$ smoothly changes and, in contrast to the lower dimensional systems, this lowest energy state is actually extended over the section in which the rotation angle is almost constant, suggesting it is not of quasi-MBS nature.

This result in combination with Secs.~\ref{Sec:3dModelSmoothDecay}  and \ref{Sec:3dModelDomainWall} suggest that the formation of trivial sub-gap states due to non-periodic rotations of the magnetic moments along the chain is unlikely in three-dimensional systems. We again note, however, that simulations in three dimensions are limited and that these results do not exclude the presence of trivial zero-energy sub-gap states, however, our results show that these states are less prevalent  in three-dimensional systems than in one- and two-dimensional systems and far less abundant than in equivalent nanowire systems.

\section{Conclusions \label{Sec:Conclusion}}
In this paper, we investigated the prevalence of zero-energy bound states due to non-periodic helical spin chains of magnetic adatoms on the surface of superconductors. We showed that there exists a mapping of the Hamiltonian of a one-dimensional helical spin chain to an effective Hamiltonian reminiscent of a Rashba nanowire in the presence of a magnetic field. This mapping shows that a spatially varying rotation rate transforms to a non-uniform Rashba SOI and a non-uniform potential in the effective chain Hamiltonian. Since these spatially varying potentials, when sufficiently smooth, are known to support the formation of trivial zero-energy sub-gap states in Rashba nanowires, the mapping therefore suggests that trivial zero-energy states in helical chains might be as abundant as in Rashba nanowires. However, unlike any realistic nanowire, the helical spin chain is installed on the surface of the superconductor. As such, although it is possible to use this mapping to generate some trivial zero-energy bound states for spin rotation profiles which mimic known mechanisms for trivial sub-gap states in Rashba nanowires, we found that such states are far less abundant in helical spin chains than in nanowires. In particular, for the most experimentally relevant setup of a helical spin chain on the surface of a three-dimensional superconductor we did not find zero-energy bound states for any rotation profile.

Although we stress that our findings do not conclusively rule out the appearance of trivial zero-energy sub-gap states in helical spin chain systems due to non-uniformities, they clearly show that the same mechanisms that lead to an abundance of zero-energy bound states in Rashba nanowires do not result in equivalent issues in atomic chains, despite an apparent mapping between the two systems. Further mechanisms other than the variation of the rate $\Phi_n$ still can result in zero-energy sub-gap states. For instance, another mechanism has been suggested in Ref.~\citenum{Sau2015Bound}: multiple YSR states, emerging from a magnetic chain on top of a superconductor, form a YSR band with van Hove singularities, visible in the LDOS. The energy of these singularities changes in the LDOS close to the chain ends and it can be tuned to zero for sufficiently strong exchange coupling.

We also want to emphasize that our findings are only relevant for long helical spin chains consisting of many rotation periods as, for instance, weakly coupled YSR states  in short trivial ferromagnetic chains can be tuned close to zero energy for certain exchange coupling strengths \cite{kuester2021nonmajorana,Howon2018Toward,schneider2021controlled,Schneider2021Topological}. Nonetheless, it is a significant benefit that zero-energy bound states can be more conclusively identified as MBSs in helical spin chains compared to in semiconductor nanowires, especially since atomic chains have a reduced tunability in comparison to semiconductor nanowire devices and so the phase space of a purported MBS signal is more difficult to explore. Our findings coupled with other benefits of helical spin chains, such as the fact that states in atomic chains can be addressed very locally via STM measurements, should enable one to build more confidence that a given zero-energy mode is of topological origin rather than trivial in nature.

\acknowledgements
This project has received funding from the European Union’s Horizon 2020 research and innovation programme under Grant Agreement No 862046 and under Grant Agreement No 757725 (the ERC Starting Grant). This work was supported by the Georg H. Endress Foundation and the Swiss National Science Foundation.

\appendix

\section{Chain on the boundary of a two-dimensional system \label{App:2dChainOnEdge}}

\begin{figure}[t]
\subfloat{\label{figTwodimModelChainOnEdge}\stackinset{l}{0.0in}{t}{0.in}{(a)}{\includegraphics[width=1\columnwidth]{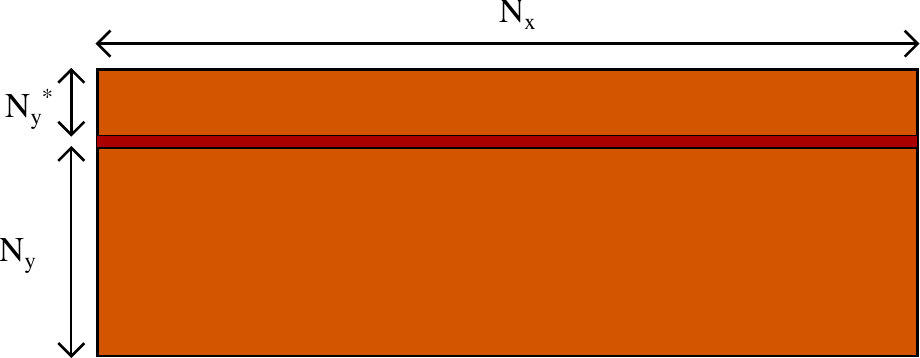}}
}\\
\subfloat{\label{fig2ChainOnTheBoundaryDependence}\stackinset{l}{0.0in}{t}{0.in}{(b)}{\stackinset{l}{1.75in}{t}{0.in}{(c)}{\stackinset{l}{0in}{t}{1.1in}{(d)}{\stackinset{l}{1.75in}{t}{1.1in}{(e)}{\includegraphics[width=1\columnwidth]{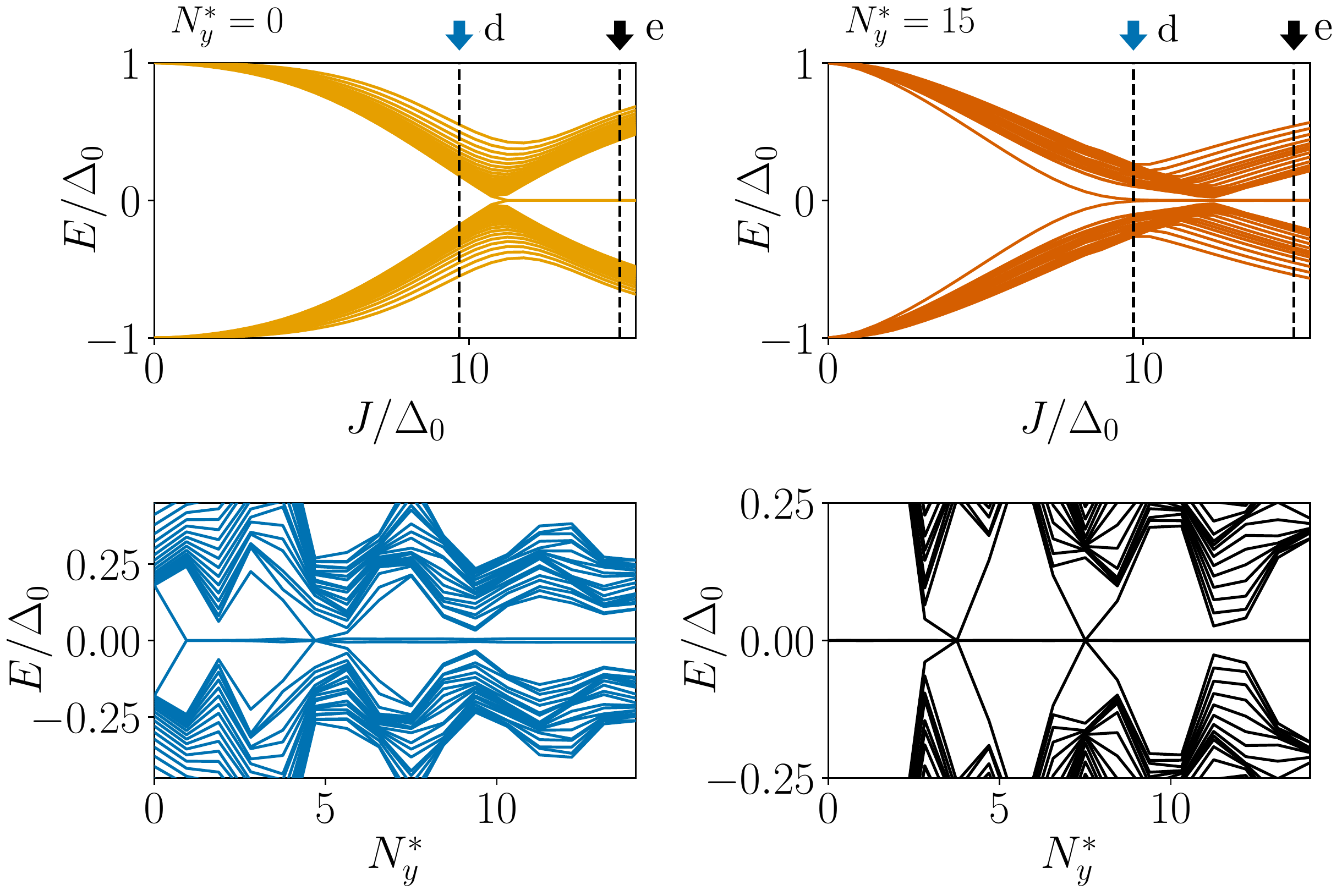}}}}}
}
\subfloat{\label{fig2dChainOnTheBoundaryLargeNyStar}}\\
\subfloat{\label{fig2dChainOnTheBoundarySmallJ}}\\
\subfloat{\label{fig2dChainOnTheBoundaryLargeJ}}\\
 \caption{ \textit{Sketch and energy spectra of a chain, with a quasi-MBS rotation profile deposited at varying distances  from the boundary in $y$ direction of a two-dimensional superconductor.}  (a) Schematic sketch of an atomic chain (red) deposited on a two-dimensional superconductor (orange). The length $N_C$ of the atomic chain coincides with the total length  $ N_x$ of the system, the chain is separated by a number of $N_y^*$ sites from the boundary of the two-dimensional superconductor. (b,c) The energy spectrum of the system as a function of the exchange coupling for $N_y^*=0$ and $N_y^*=15$, respectively.  (d,e) Energies as a function of $N_y^*$ for an exchange coupling smaller  and larger than the critical exchange coupling. The  blue [black] arrow in panels (a) and (b) indicates the value of the exchange coupling used in panel (d) [(e)]. The trivial zero-energy state is present only if the chain is placed away for the boundary.  The parameters are the same as in Fig.~\ref{figHelicalChainDirectVsIndirectDependenceInTwoDimensionsSQuasiMBSs}, except that we introduced the quantity $N_y^*$.}
 \label{fig2dChainOnEdge}
\end{figure}

In this Appendix, we vary the position $m_0$ of the atomic chain in two-dimensional systems. In particular, we place the chain on a boundary of a two-dimensional superconductor. This analysis is motivated by our finding in three-dimensional systems (see Secs.~\ref{Sec:3dModelSmoothDecay}, \ref{Sec:3dModelDomainWall}, and \ref{Sec:3dModelQuasiMBSs}) that the appearance of trivial zero-energy sub-gap states is suppressed compared to similar one- and two-dimensional systems where such states occur due to smooth variations of the rotation rates of neighbouring magnetic moments along the atomic chain. This effect is likely a result of the boundary conditions, since the atomic chain is placed on the surface of the three-dimensional superconductor. In the two-dimensional system, see Secs.~\ref{Sec:2dModelSmoothDecay}, \ref{Sec:2dModelDomainWall}, and \ref{Sec:2dModelQuasiMBSs}, however, we considered  a chain in the bulk of the superconductor ($m_0 \approx N_y/2$). Placing the chain on the boundary of a two-dimensional superconductor should, therefore, lead to similar boundary effects as observed in three dimensions. 

To further investigate the importance of boundary effects, we consider the model from Sec.~\ref{Sec:2dModelQuasiMBSs}, with identical parameters except that the chain is located at the boundary of the two-dimensional superconductor, $m_0=0$. The corresponding energy spectrum as a function of the exchange coupling $J$ is shown in Fig.~\ref{fig2ChainOnTheBoundaryDependence}. The bulk gap closing and reopening at a critical exchange coupling is accompanied by the appearance of MBSs at the ends of the chain. The system does not, however, host trivial zero-energy sub-gap states for $J$ smaller than the critical exchange coupling. 

Next, we extend the superconductor in $y$ direction, by adding $N_y^*N_x$ sites, so that the atomic chain is not anymore positioned  on the boundary of the two-dimensional system. We calculate the energies as a function of $N_y^*$ for an exchange coupling smaller (see Fig.~\ref{fig2dChainOnTheBoundarySmallJ}) and larger (see Fig.~\ref{fig2dChainOnTheBoundaryLargeJ}) than the coupling necessary for the bulk gap closing and reopening.  
In the first case, there are no zero-energy states for $N_y^*=0$, however, for finite values of $N_y^*$ a state with almost zero energy appears.  Furthermore, the size of the superconducting gap changes, due to varying leakage of the wavefunctions to sites neighbouring the atomic chain. For sufficiently large values of $N_y^*$ the gap and the zero-energy state stabilize. In contrast, when the entire atomic chain enters the topological regime, MBSs appear also  for the choice $N_y^*=0$ and are stable against changes of $N_y^*$.  In both cases the size of the superconducting gap varies as a function of $ N_y^*$, but it stabilizes for sufficient large values of $N_y^*$.

Finally, we plot the energy spectrum as a function of the exchange coupling $J$ in the limit of large $N_y^*$, which means that the localization length of the states in $y$ direction is shorter than the length $aN_y^*$, see Fig.~\ref{fig2dChainOnTheBoundaryLargeNyStar}.  For this case, we obtain almost the same spectrum as in Fig.~\ref{figEnergiesAligned2dQuasiMBS}, as expected. In addition, we calculated the energies of the systems from Secs.~\ref{Sec:2dModelSmoothDecay} and \ref{Sec:2dModelDomainWall} with $m_0=0$. In both cases, we did not find any trivial zero-energy sub-gap states in agreement with the results presented in Fig.~\ref{fig2dChainOnEdge}. In conclusion, we find that placing the chain at the boundary of the superconductor, such that it is parallel to the chain, suppresses the formation of trivial zero-energy sub-gap states in two dimensions.

\section{Chain in the bulk of a three-dimensional superconductor \label{App:ChainIn3dBulk}}
 \begin{figure}[!b]
\subfloat{\label{fig3D_QuasiMBSsProfile_EnergySpec_ChainInBulk}\stackinset{l}{0.in}{t}{0in}{(a)}{\stackinset{l}{0.in}{t}{1.25in}{(b)}{\stackinset{l}{0.2in}{t}{-1.1in}{\includegraphics[width=0.7\columnwidth]{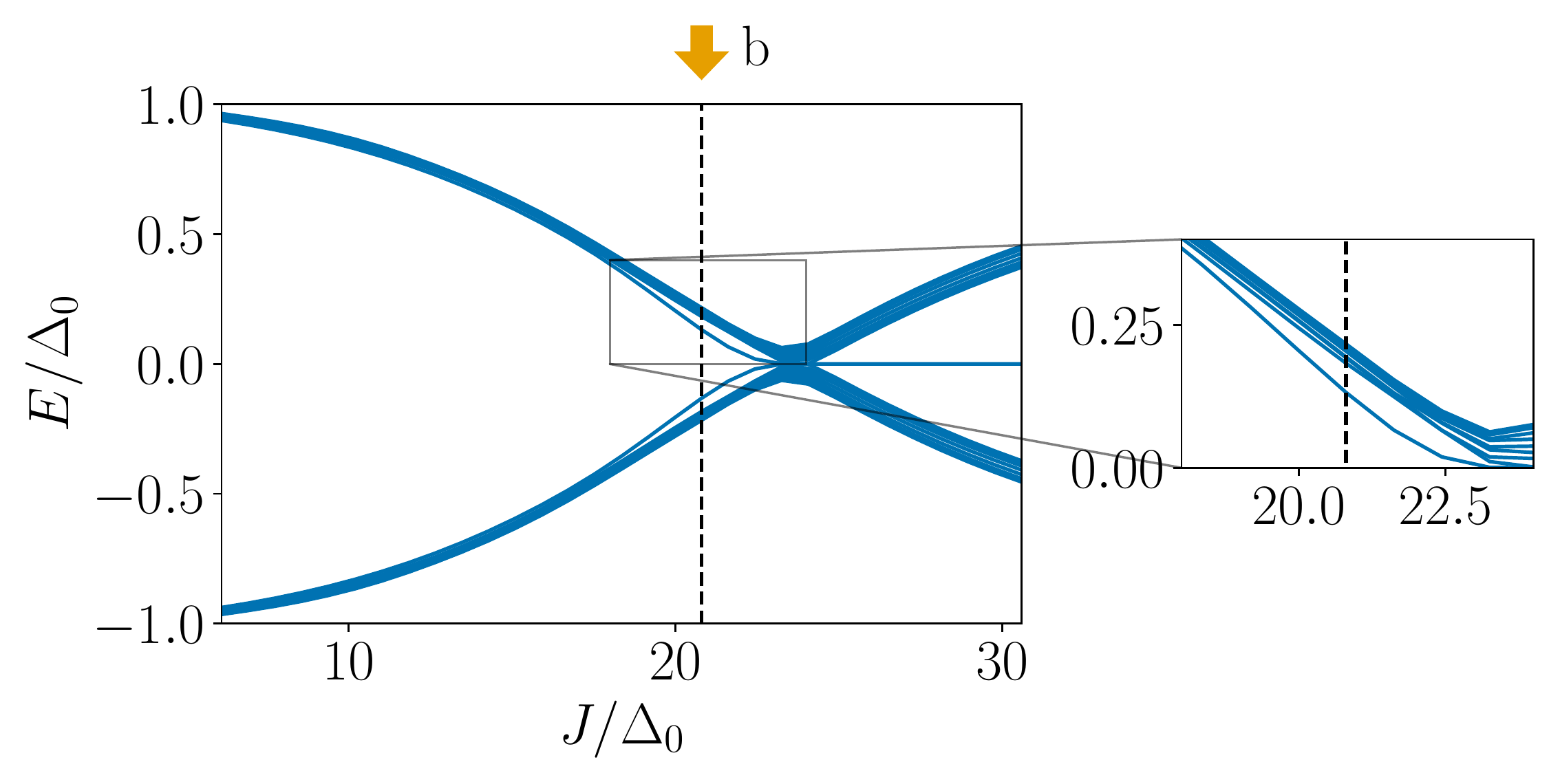}}{\includegraphics[width=1\columnwidth]{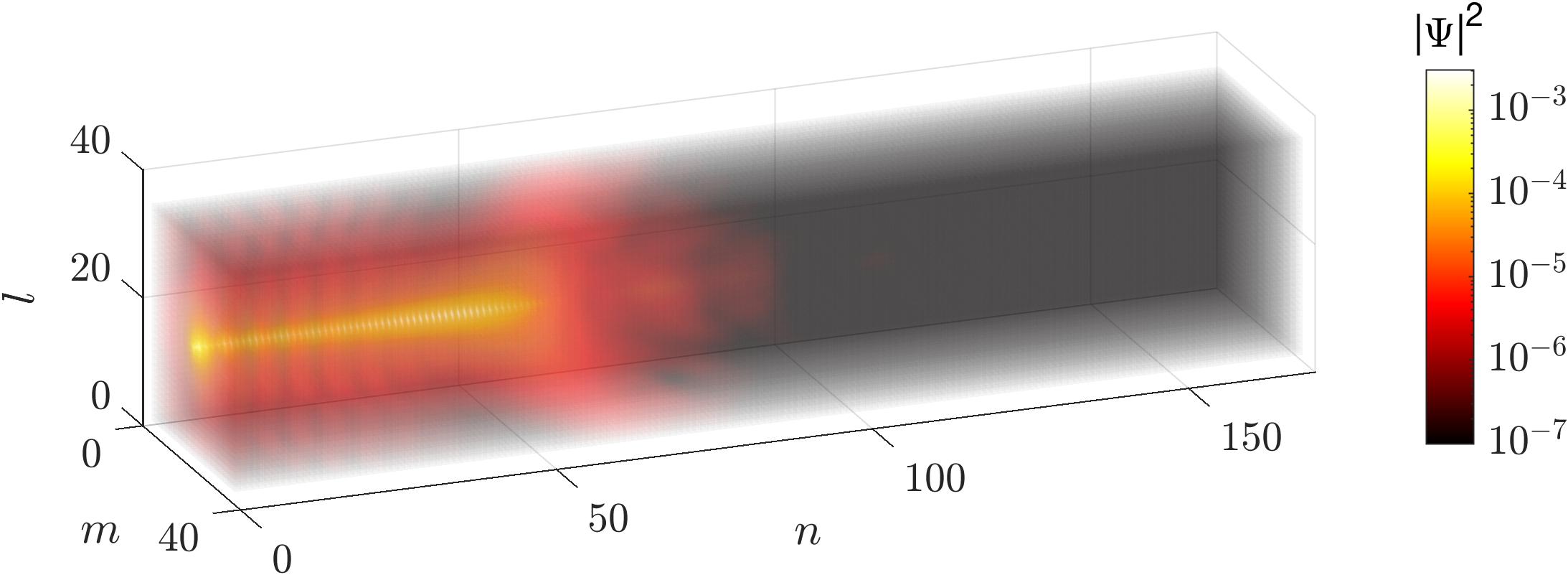}}}}
}
\subfloat{\label{figProbDens3d_QuasiMBS_Triv_ChainInBulk}}\\
 \caption{\textit{Energy spectrum and probability density of the lowest energy state of a chain with a $\underline{\text{quasi-MBS}}$ rotation profile deposited in the $\underline{\text{bulk}}$ of a $\underline{\text{three}}$-dimensional superconductor.}  (a) The system hosts sub-gap states and a zoom into the figure reveals a clear separation of a sub-gap state from the bulk states. (b) Probability density of the sub-gap state at the exchange coupling indicated by the yellow arrow in panel (a).  The trivial sub-gap state is localized in the region of smooth change of the rotation rate. The only changed parameters compared to Fig.~\ref{fig3dQuasiMBSs} are the locations of the chain sites, such that $N_y=N_z=35$ and $m_0=l_0=17$.}
 \label{figChainIn3dBulk}
\end{figure}

In experiments, see e.g. Refs.~\cite{Menzel2012Information,Howon2018Toward,Schneider2021Topological}, magnetic adatoms are deposited on the surface of a superconductor, therefore we consider in the main text only chains which are placed  on the surface of a three-dimensional superconductor. Here, on the contrary, we study a chain {\it located fully inside the bulk} of a three-dimensional system. This particular situation cannot easily be realized in experiments, however, we study this scenario to gain further insights about the importance of boundary effects on the sub-gap state at the surface. In particular, a sub-gap state appears in this setup for exchange couplings smaller than the critical value at which the gap closes and reopens, see Fig.~\ref{fig3D_QuasiMBSsProfile_EnergySpec_ChainInBulk}. This sub-gap state is not pinned to zero energy, but it is separated in energy from the bulk gap, see the zoom on the right side,  which is not the case for systems in which the chain is placed on the surface of the superconductor, cf.  Secs.~\ref{Sec:3dModelSmoothDecay}, \ref{Sec:3dModelDomainWall}, and \ref{Sec:3dModelQuasiMBSs}. Furthermore, we note that the energy of the sub-gap state could potentially be pinned to zero for sufficiently large systems. Moreover, the  state is localized in the region of  smooth change of $\Phi$, see Fig.~\ref{figProbDens3d_QuasiMBS_Triv_ChainInBulk}. These results in combination with Appendix~\ref{App:2dChainOnEdge} clearly show that the relative position of the chain towards the superconductor boundary can strongly affect the energies of sub-gap states and, in particular, trivial zero-energy states are suppressed when the chain is placed close to a boundary, such as the surface of a three-dimensional superconductor.

\bibliographystyle{apsrev4-1}
\bibliography{Literatur3}
\end{document}